\documentclass[11pt,a4paper]{article}

\usepackage[T1]{fontenc}
\usepackage[utf8]{inputenc}
\usepackage{lmodern}
\usepackage[margin=2.5cm]{geometry}
\usepackage{amsmath,amssymb,bm}
\usepackage{graphicx}
\usepackage{xcolor}
\usepackage{booktabs}
\usepackage{array}
\usepackage{textcomp}
\usepackage{placeins}
\usepackage{caption}
\usepackage{newfloat}
\usepackage[numbers,sort&compress]{natbib}
\usepackage[colorlinks=true,linkcolor=blue!50!black,citecolor=blue!50!black,urlcolor=blue!50!black]{hyperref}

\DeclareFloatingEnvironment[fileext=loe,listname={List of Extended Data Figures},name={Extended Data Figure},placement=tbp]{edfigure}
\DeclareFloatingEnvironment[fileext=los,listname={List of Supplementary Figures},name={Supplementary Figure},placement=tbp]{sifigure}
\newcommand{\figbox}[4][0pt 0pt 0pt 0pt]{
  \begin{center}
  \IfFileExists{figures/#2.pdf}{\includegraphics[width=\linewidth,trim=#1,clip]{figures/#2.pdf}}{%
  \IfFileExists{figures/#2.png}{\includegraphics[width=\linewidth,trim=#1,clip]{figures/#2.png}}{%
  \IfFileExists{figures/#2.jpg}{\includegraphics[width=\linewidth,trim=#1,clip]{figures/#2.jpg}}{%
    \fbox{\parbox[c][#3][c]{0.96\linewidth}{\centering\textbf{FIGURE PLACEHOLDER}\\[3pt]
    \texttt{figures/#2.pdf}\\[3pt]\small #4}}}}}
  \end{center}}

\newcommand{\Rsun}{\ensuremath{R_{\odot}}}
\newcommand{\kms}{\,km\,s\ensuremath{^{-1}}}
\newcommand{\Wm}{\,W\,m\ensuremath{^{-2}}}
\newcommand{\ergcgs}{erg\,cm\ensuremath{^{-2}}\,s\ensuremath{^{-1}}}
\newcommand{\mHz}{\,mHz}
\newcommand{\arcsec}{\ensuremath{^{\prime\prime}}}
\newcommand{\FeIX}{Fe\,\textsc{ix}}
\newcommand{\FeXIII}{Fe\,\textsc{xiii}}

\newcommand{\dd}{\mathrm{d}}

\title{A magneto-convective origin for solar-wind Alfvénic fluctuations}

\author{R.~J.~Morton$^{1,*}$, Z.~Huang$^{2}$, T.~Schad$^{3}$, M.~Velli$^{2}$\\[4pt]
{\small $^{1}$School of Engineering, Physics and Mathematics, Northumbria University, Newcastle, UK}\\
{\small $^{2}$Department of Earth, Planetary, and Space Sciences, University of California, Los Angeles, California, 90095, USA}\\
{\small $^{3}$National Solar Observatory, Makawao, Hawaii, 96768, USA}\\
{\small $^{*}$Corresponding author: \href{mailto:richard.morton@northumbria.ac.uk}{richard.morton@northumbria.ac.uk}}}
\date{}

\begin{document}
\maketitle

\begin{abstract}
The solar wind is a highly structured and supersonic plasma outflow that originates in the Sun’s outer atmosphere \cite{Neugebauer1962Science,DeForest2018ApJ}. The energy source driving the wind is still subject to intense debate. Recent observations point to magnetic reconnection as a primary driver \cite{Bale2023Nature,Chitta2023Science,Raouafi2023ApJ}, while Alfvénic waves launched by surface convection are often invoked as a key contributor \cite{Suzuki2005ApJL,Cranmer2007ApJS}. Whether fluctuations in the young solar wind originate in convectively driven waves has not been tested observationally. Here we show that they carry the spectral imprint of convective driving. Observations from the Solar Dynamics Observatory \cite{Pesnell2012SoPh} and the Upgraded Coronal Multi-channel Polarimeter \cite{Landi2016JGR} reveal the predicted spectrum of convectively driven waves in both weakly and rapidly expanding open fields, the likely sources of the fast and slow Alfvénic wind. Averaged over 270 intervals below the Alfvén surface, the young wind sampled by Parker Solar Probe \cite{Fox2016SSRv} carries a similar spectrum. High-resolution spectroscopy from the Daniel K. Inouye Solar Telescope \cite{Rimmele2020SoPh} resolves further spectral components that also appear in the wind. Convectively driven waves therefore supply a basal energy flux to the solar wind, alongside intermittent reconnection.
\end{abstract}

\section*{Main Article}

Both the ubiquity of Alfvénic fluctuations and their role in energy transfer throughout the Sun’s chromosphere and corona are well established \cite{Tomczyk2007Science,DePontieu2007Science,McIntosh2011Nature,Morton2019NatAstron}. In coronal holes, the source regions of the fast solar wind, measurements from across a solar cycle demonstrate that velocity amplitudes associated with such fluctuations are consistently around 20--30\kms\ at 1.03\,\Rsun\ \cite{Morton2025ApJ985}, which translates to a typical wave energy flux of $\sim$100\Wm. The recent identification of small-scale torsional Alfvénic modes in the corona could double this estimate for energy flux \cite{Morton2026NatAstron}. Comparable wave fluxes are also reported in other open field regions and the quiet Sun \cite{McIntosh2011Nature,Morton2019NatAstron}. The origin of such waves has long been proposed to be the Sun’s turbulent convective motions at the surface \cite{Alfven1947MNRAS,Piddington1956MNRAS}. These motions can shake and twist the base of the magnetic flux tubes \cite{Bonet2008ApJL,Chitta2012ApJ} that reach upwards through the Sun’s atmosphere and eventually expand to form the open flux in the heliosphere. Many numerical simulations have demonstrated that the upward propagation of Alfvénic waves from the photosphere and their subsequent dissipation can lead to hot, fast outflows with physical properties comparable to the fast wind streams \cite{Suzuki2005ApJL,Cranmer2007ApJS}.

In contrast, several recent studies \cite{Bale2023Nature,Chitta2023Science,Raouafi2023ApJ} have emphasised the role of interchange reconnection (i.e., between open and closed magnetic field lines), suggesting it is the dominant driver of the solar wind. Central to this idea is the observed dominance of switchbacks and switchback patches in the young solar wind. Switchbacks are spherically polarised Alfvénic fluctuations of sufficient amplitude to reverse the magnetic field, and, consequently, carry outward jets with magnitude comparable to the Alfvén speed \cite{Horbury2023PhPl}. The switchbacks are estimated to carry significant momentum and energy for driving the solar wind. Results from the Parker Solar Probe (PSP) indicate that the switchbacks appear in patches whose spatio-temporal structure is well correlated with the scales of the Sun's surface super-granulation \cite{Bale2021ApJ}. Supergranulation also modulates the underlying base solar wind speed. These findings suggest a low-coronal interchange-reconnection origin for switchbacks and may also help explain associated in-situ ion spectra \cite{Bale2023Nature}. Inner coronal observations have added support to this by revealing the presence of small-scale jet phenomena arising from reconnection between small-scale opposite polarity magnetic flux \cite{Raouafi2014ApJ,Chitta2023Science,Raouafi2023ApJ}.

However, this picture is incomplete. Modelling shows that switchbacks resulting from reconnection tend to relax and the field tends to straighten out extremely rapidly, below where PSP measures them \cite{Landi2005ESASP,Uritsky2017ApJ} ($<$10\,\Rsun). Indeed, more promising formation mechanisms have switchbacks emerging from the background Alfvénic turbulence itself, either as a result of the coronal and solar wind expansion \cite{Squire2020ApJL,Shoda2021ApJ} or the interaction with transverse shears in the phase speed \cite{Toth2023ApJ}. Also, while reconnection may provide a significant fraction of the heating of the corona, interchange reconnection alone may not explain the fluctuations and energy budget of the solar wind. While the modulation of the baseline wind speed in patches could be a coronal heating effect, it might be unrelated to the velocity jets of individual switchbacks in patches, resulting instead from the properties of spherically polarised Alfvénic fluctuations. Also, Eugene Parker himself \cite{Parker1991ApJ} noted how the fast solar wind required an addition of energy of order 100\Wm\ to boost the wind from an already heated coronal hole up to asymptotic wind speeds above 700\kms.

\begin{figure}[t]
\figbox[0 0 0 0]{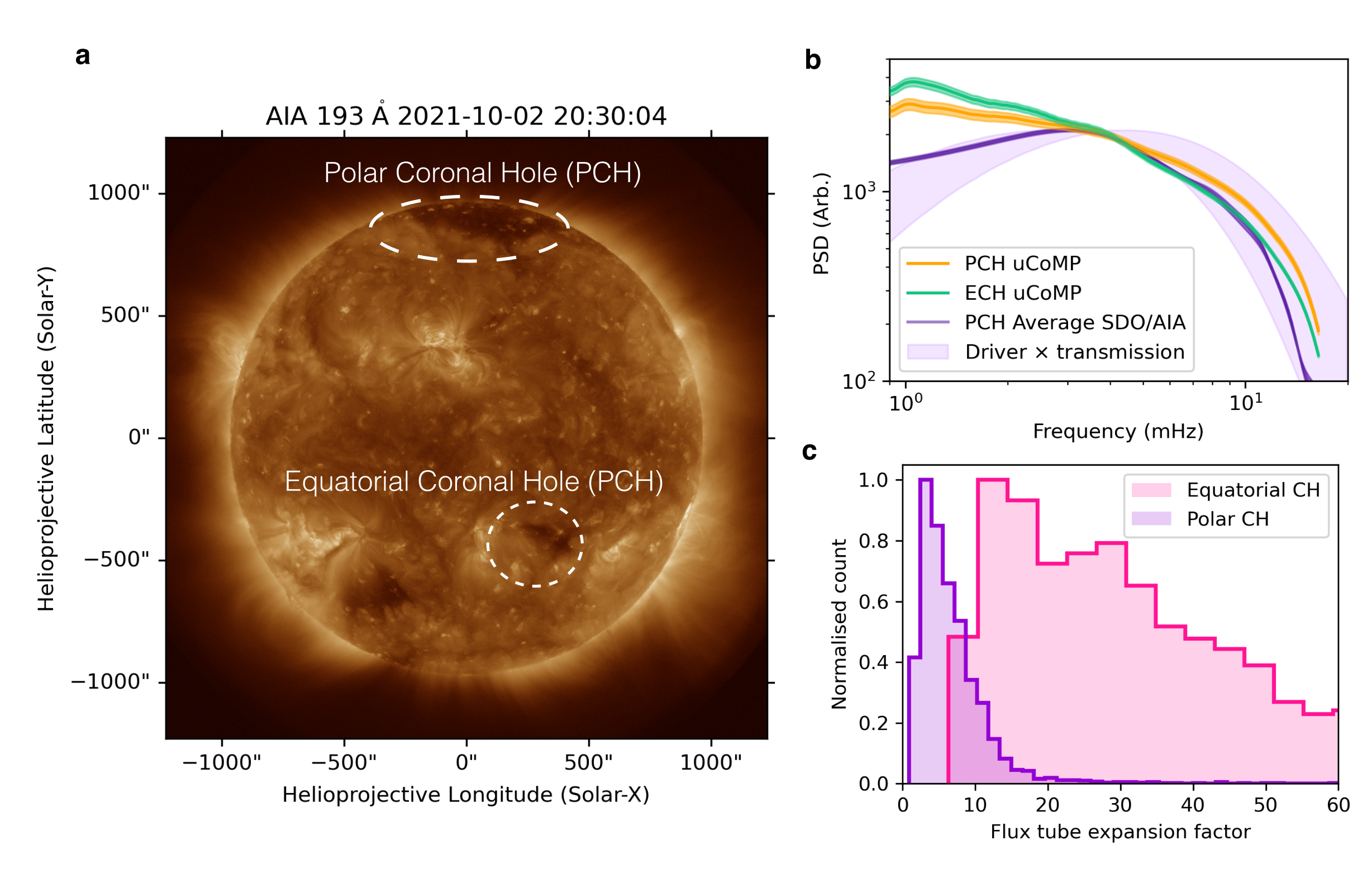}{9cm}{Panels (a)--(c): SDO/AIA 19.3\,nm context image with polar and equatorial coronal holes (a); UCoMP and SDO/AIA Doppler-velocity PSDs vs.\ theoretical transmitted spectrum (b); magnetic expansion factors (c).}
\caption{\textbf{The solar wind originates in polar and equatorial coronal holes} (panel a). The image shows the corona as observed with SDO/AIA (19.3\,nm) on the 2 October 2021, three days before the UCoMP observations, when the equatorial hole was on disk. The Doppler velocity power spectral density from UCoMP for the Alfvénic fluctuations within the polar (PCH, yellow) and equatorial (ECH, green) coronal holes is shown in panel b. Also displayed is the average power spectral density (purple) of Alfvénic fluctuations obtained from SDO/AIA data, averaged across 8 polar coronal holes at different points in the solar cycle. For both UCoMP and SDO/AIA measurements, the thin curve represents the mean value and the lighter shaded band represents the 95\% confidence interval on mean. The expected profile of Alfvénic fluctuations entering the corona based on a theoretical transmission profile and a measured photospheric transverse driving spectrum is shown in lilac. The width of the curves represents a range based on different photospheric magnetic field strengths. Panel c shows the estimated expansion for the coronal holes present in panel a, calculated between the photosphere and the source surface.}
\label{fig:overview}
\end{figure}

If convection drives the Alfvénic fluctuations, a specific fossil spectrum should be present in the open corona regardless of magnetic field expansion in the corona and persist until at least Alfvén surface \cite{Cranmer2007ApJS,Cranmer2005ApJS}. To assess the potential contribution of convectively-driven coronal Alfvénic fluctuations to the solar wind, we examine the properties of the waves across multiple sets of inner coronal observations. The targets are the weakly expanding magnetic fields in the cores of polar coronal holes (and source regions of the fast wind) and the rapidly expanding magnetic fields of an equatorial coronal hole (suspected sources of the slow Alfvénic wind), with examples shown in Figure~\ref{fig:overview}a. Using imaging data from the 17.1\,nm (\FeIX) channel of the Atmospheric Imaging Assembly (AIA) \cite{Lemen2012SoPh} on the Solar Dynamics Observatory (SDO), we analysed Alfvénic motions in polar coronal holes across the solar cycle (see Methods and Extended Data Figure~\ref{ed:aia}). The results reveal that the power spectral density (PSD), $P(f)$, shows a broad enhancement that peaks around 4\mHz\ (Figure~\ref{fig:overview}b). The shape of the power spectra is similar to previous Doppler velocity measurements in \FeXIII\ observed throughout the rest of the corona \cite{Morton2019NatAstron} but were signal-to-noise limited in the cores of coronal holes (e.g., Extended Data Figure~\ref{ed:ucomp}). The PSD closely agrees with a theoretically predicted (see Methods) range for coronal Alfvénic wave spectrum from magneto-convective driving (Figure~\ref{fig:overview}b). The distinct shape arises due to reflection of low-frequency Alfvénic waves at the Transition Region and damping of high frequency waves in the chromosphere from phase mixing and ambipolar diffusion \cite{Morton2025ApJL,Soler2017ApJ,Soler2019ApJ}.

The Upgraded Coronal Multi-Channel Polarimeter \cite{Landi2016JGR} (UCoMP) provides spectro-polarimetric coronagraphic observations of the 1074\,nm \FeXIII\ emission line, enabling direct measurements of coronal Doppler velocity fluctuations, which are known to be dominated by Alfvénic fluctuations \cite{Tomczyk2007Science,Morton2019NatAstron,Morton2025ApJ985}. For 5 October 2021, both polar and equatorial coronal holes are present in the off-limb data (Figure~\ref{fig:overview}a shows the corona 3 days prior when the equatorial hole is on disk). Due to the weak \FeXIII\ emission, there is only meaningful signal-to-noise at the base and edge of the polar hole (Extended Data Figure~\ref{ed:ucomp}). Magnetic field expansion rates (Figure~\ref{fig:overview}c) from potential field extrapolations (see Methods and Extended Data Figure~\ref{ed:pfss}) demonstrate that the northern polar and equatorial coronal holes have the desired geometry to be potential fast and slow Alfvénic wind sources, respectively. The Doppler velocity PSD for these two regions are comparable to the PSDs measured with SDO/AIA. The UCoMP spectra deviate from the SDO/AIA spectra at low frequencies due to the dominance of a power-law component in this range \cite{Morton2019NatAstron}. The analysis of the SDO/AIA imaging data only counts identified oscillatory events, so the background power law never enters the AIA spectrum, whereas it does in a Doppler PSD. The convective origin of coronal waves is further supported by previous measurements of the perpendicular correlation length scales associated with transverse motions in the photosphere \cite{Abramenko2013ApJ,Chhiber2025MNRAS} ($\sim$1.5\,Mm) and corona \cite{Sharma2023NatAstron} ($\sim$8\,Mm), which are consistent assuming that the correlation length scale varies in proportion to the expansion of the magnetic field with radial distance \cite{Hollweg1986JGR}. Hence, the results indicate that Alfvénic waves in open-field regions, at the footpoints of both slow and fast wind streams, are driven primarily by surface convection.

\begin{figure}[t]
\figbox[0 0 0 0]{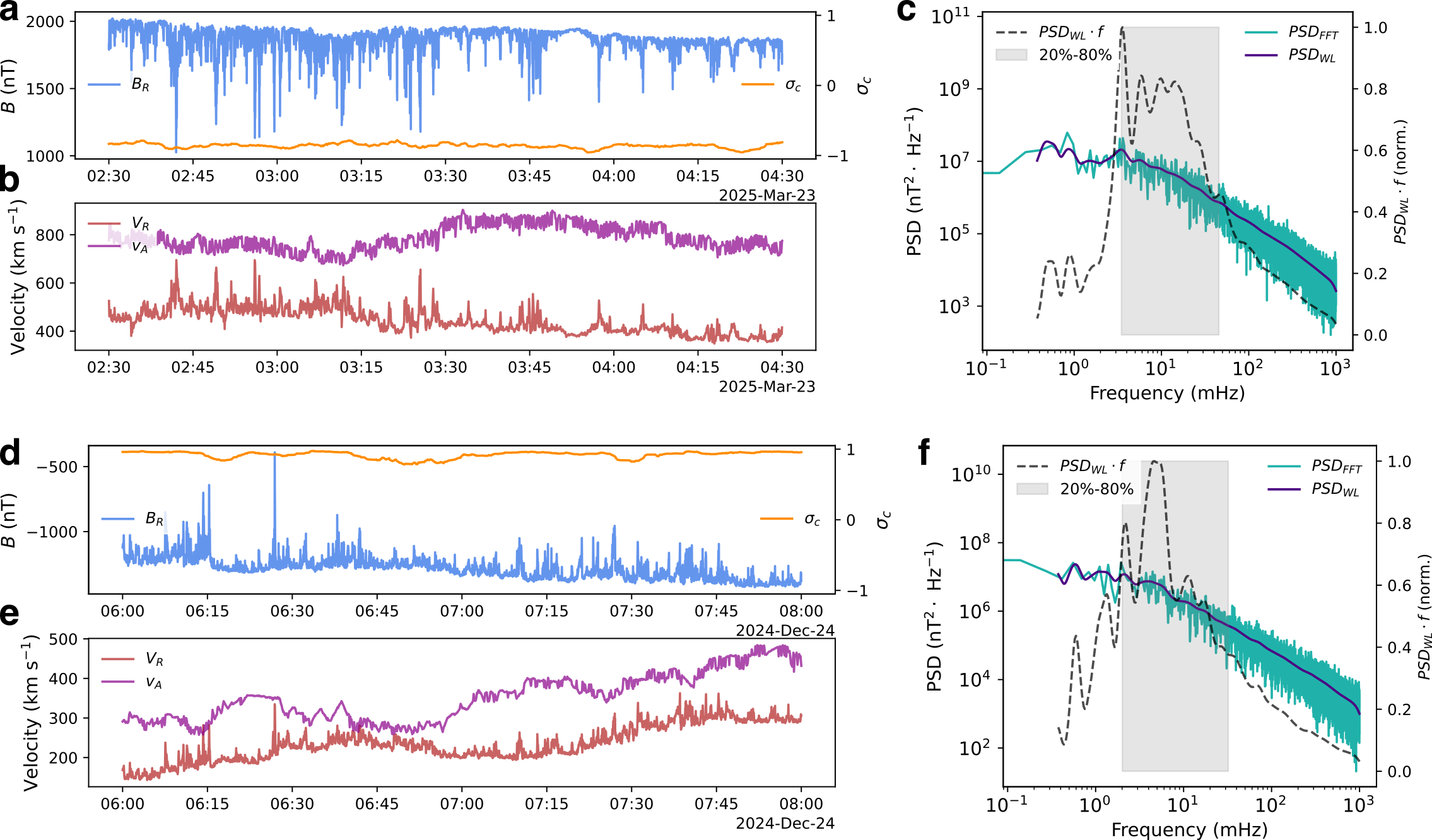}{10cm}{Panels (a)--(f): PSP fast-wind (a--c) and slow-wind (d--f) streams: radial magnetic field and cross-helicity; V$_R$ and V$_A$; trace magnetic PSDs (Fourier and wavelet) and energy density P(f)$\cdot$f.}
\caption{\textbf{Fluctuations in slow and fast wind streams.} Panels a--c show PSP data from the fast wind, while panels d to f are related to the slow stream. Panel a(d) shows the radial component of the magnetic field ($B_{R}$), which shows magnetic field deflections (short-scale, intermittent changes) often associated with switchbacks. Also shown is the cross-helicity, that signifies the fluctuations are Alfvénic in nature. Panel b(e) displays the radial velocity ($V_{R}$), which is solar wind outflow speed and the Alfvén speed ($V_{A}$). Panel c(f) contains the trace power spectral density (Fourier, $\mathrm{PSD}_{\mathrm{FFT}}$ and wavelet, $\mathrm{PSD}_{\mathrm{WL}}$) for the magnetic field. The grey shaded region shows the frequency interval that contains the middle 60\% of the fluctuation energy. The dashed lines are the wavelet energy density, $\mathrm{PSD}_{\mathrm{WL}}\cdot f$. Both the shaded region and the energy density indicate that the fluctuation energy in the solar wind is concentrated around 1--30\mHz\ in both fast and slow streams.}
\label{fig:psp}
\end{figure}

\begin{figure}[t]
\figbox[0 0 0 0]{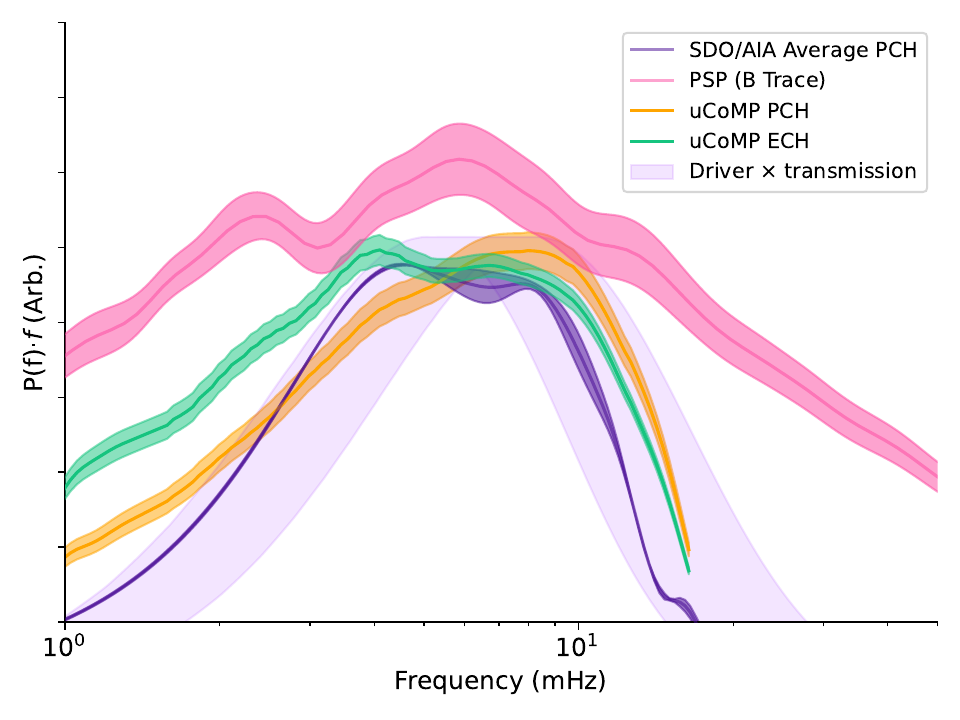}{9cm}{Energy density per logarithmic frequency interval vs.\ frequency (mHz): UCoMP (PCH, ECH), SDO/AIA, PSP, and the theoretical transmission$\times$driver band (800--1500\,G).}
\caption{\textbf{A comparison of the fluctuation energy density per logarithmic frequency interval} for measurements of Alfvénic fluctuations from UCoMP (yellow and green), SDO/AIA (purple) and PSP (pink). For the observational measurements the thin curve represents the mean value and the lighter shaded band represents the 95\% confidence interval on mean. The light purple shaded region is the expected energy density estimated by combining a theoretical transmission profile with the observed power spectrum from photospheric bright point motions. The shaded region covers transmission profiles associated with photospheric field strengths from 800--1500\,G. The dark purple curve indicates the mean energy density estimated from all the individual Alfvénic wave velocity power spectra within the polar coronal holes observed with SDO/AIA. The pink curve is the mean PSP magnetic trace wavelet energy density averaged over all Alfvénic intervals below the Alfvén surface from perihelion during encounters 10--24. Note the scaling between the data sets is arbitrary, partly due to issues with estimating the magnitude of the PSD from the SDO data.}
\label{fig:compare}
\end{figure}

PSP measures fluctuations in the solar wind plasma in situ. Since Encounter 8, PSP's perihelia have taken it below the Alfvén surface and into the young solar wind (Extended Data Figure~\ref{ed:psp}i). We selected all intervals from Encounters 10--24 that lie within $15\,\Rsun$, below the Alfvén surface, and have a large cross helicity ($\vert \sigma_{c}\vert > 0.8$), giving 270 two-hour windows of highly Alfvénic wind (Methods). Examples from a fast and a slow stream (Figure~\ref{fig:psp}) show the hallmarks of Alfvénic turbulence: large-amplitude, intermittent deflections of the radial magnetic field (Figure~\ref{fig:psp}a,d) accompanied by high cross helicity, which remains high across the energy-containing frequencies (e.g., Extended Data Figure~\ref{ed:psp}b). The trace magnetic-field power spectral densities (Figure~\ref{fig:psp}c,f) show a shallow power law at low frequencies, associated with the energy-injection scales, and a steeper power law at higher frequencies marking the inertial range. In both streams, the energy density per logarithmic frequency, $P(f)\cdot f$, is concentrated at a few millihertz (Figure~\ref{fig:psp}c,f). A concentration near 3--8\mHz\ has been reported in individual highly Alfvénic streams \cite{Davis2023ApJ,Huang2023ApJL,Huang2024ApJL,Huang2026ApJL}, but its origin has not been established. Averaging over the full ensemble removes the stream-to-stream variability and reveals a well-defined energy-containing band between 1--30\mHz, peaking at 5.6\mHz\ (Figure~\ref{fig:compare}). Near perihelion, PSP's radial velocity is small compared with the propagation speed of the fluctuations, so the observed frequencies differ from their values in the solar frame by less than $\sim$10\% (median $\sim$6\%), and the spacecraft's tangential motion is unlikely to alias spatial structure into the band (Methods). This holds when wave reflection is included \cite{Heinemann1980JGR,Velli1993AA}. The frequencies measured by PSP can therefore be compared directly with those in the low corona.

A direct comparison of the energy density per logarithmic frequency of the fluctuations found in the inner corona and the young solar wind near the Alfvén surface highlights they have similar distributions, peaking in the 4--8\mHz\ frequency range (Figure~\ref{fig:compare}). Table~\ref{tab:stats} quantifies this similarity giving the peak, median frequency and the 20--80\% energy interval for each dataset. Alfvénic fluctuations in this frequency range are expected to be in energy equipartition (kinetic to magnetic energy ratio of unity), an MHD-like regime, throughout the corona and solar wind \cite{Cranmer2005ApJS}. Hence the shapes of the velocity and magnetic power spectral densities can be directly compared (the absolute values cannot due to the issues with reliably measuring the magnitude of the coronal velocity power, see Methods). The coronal and solar wind energy densities align with the expected distribution from the theoretical estimates of convective wave driving (Figure~\ref{fig:compare}). There are several inferences that can be drawn from this similarity.

The first is that the Alfvénic waves observed in the corona propagate out to the Alfvén surface with the spectral content of the 1--30\mHz\ range largely intact, before being cannibalised by the turbulent cascade as the inertial range moves to lower frequencies with increasing distance \cite{Davis2023ApJ,Huang2023ApJL,Bruno2013LRSP,Chen2020ApJS,Sioulas2023ApJL}. This result is consistent with turbulence modelling \cite{Chandran2018JPP} and statistical analysis \cite{Huang2024ApJL}, which indicate several hours are required for nonlinear interactions to modify the waves’ power spectrum. The propagation time for Alfvénic waves from the coronal base to the Alfvén surface is likely under two hours \cite{Huang2024ApJL}. Therefore, the power spectrum at PSP reported here is only minimally influenced by nonlinear interactions and should be comparable to the wave power spectrum at the base of the corona.

More significantly, the correspondence between the observed energy densities and the theoretical profile would indicate that convective motions in the photosphere are a significant energy source for solar wind fluctuations. The lower solar atmosphere shapes the frequency dependence of the power spectrum for upwardly propagating Alfvénic waves, and this imprint persists as the corona transitions to the solar wind. The observed correspondence aligns with theoretical expectations, where solutions to the linear non-WKB wave transport equations show the ‘fossil’ spectrum from the Sun’s convection survives into the solar wind \cite{Cranmer2005ApJS}. We note that, when considering the full frequency scale, the frequency range for the energy density in PSP is generally broader than the theoretical coronal energy density and that of the coronal observations (Figure~\ref{fig:compare}). The broadening on the higher frequency side could be a signature of the development of turbulence between the low corona and solar wind. The broadening of the low frequency side could be other nonlinear processes, e.g., an inverse cascade, that eventually leads to the $1/f$ range formed above the Alfvén surface \cite{Huang2025ApJL}.

Although the coronal and solar-wind spectra follow the predicted transmitted spectrum, they are not fully described by it, with excess power near 4\mHz\ and 6--8\mHz\ (shown most clearly in Figure~\ref{fig:compare}). Identifying the sources of this excess requires separating overlapping spectral components, which in turn requires Doppler velocity time series with higher cadence, spatial resolution and signal-to-noise than are available from UCoMP. The Cryogenic Near Infrared Spectropolarimeter \cite{Fehlmann2023SoPh} (Cryo-NIRSP) (at the Daniel K. Inouye Solar Telescope -- DKIST \cite{Rimmele2020SoPh}) observes the same \FeXIII\ 1074\,nm line as UCoMP, at $\sim$1\,s cadence and $\sim$0.4\arcsec\ resolution.

\begin{figure}[t]
\figbox[0 0 0 0]{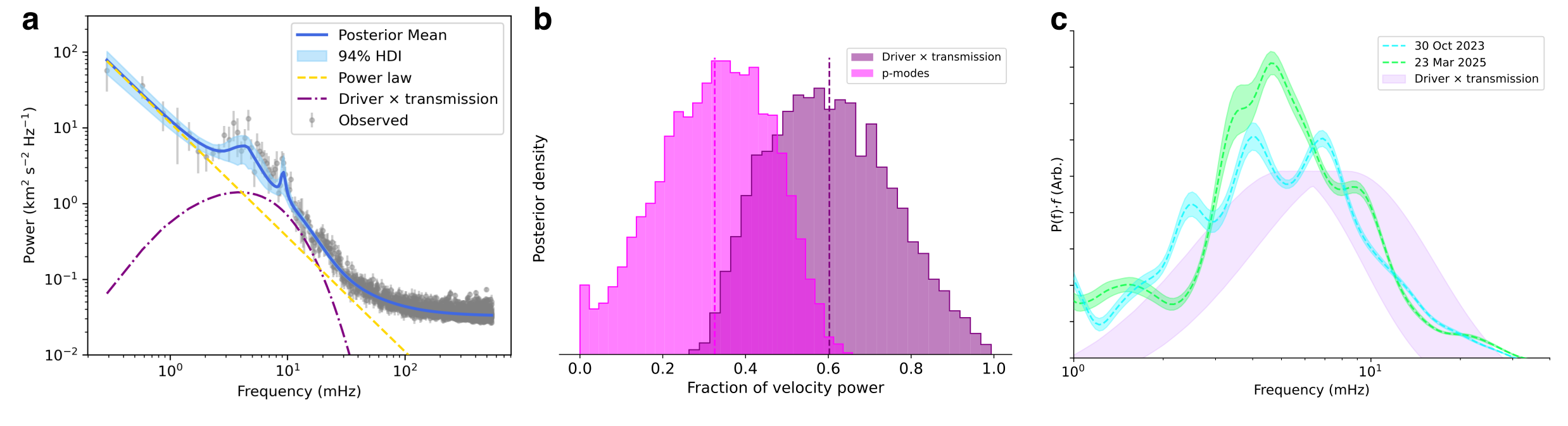}{9cm}{Panels (a)--(c): Cryo-NIRSP Doppler-velocity power spectrum with Bayesian model decomposition (a); fractional contribution of model components, 3--30\,mHz (b); energy density vs.\ theoretical coronal energy density (c).}
\caption{Panel a) shows an example Doppler velocity power spectrum (data points) estimated from the sit-and-stare observation sequence taken on 23-03-2025. The posterior mean model is shown as a blue solid line, while the corresponding highest density interval is shown as the blue shaded region. The estimated power law component (yellow) and the contribution from granulation driven, upwardly propagating Alfvénic waves (purple) are also shown. Panel b) shows the estimated contribution from granulation driven waves and that from mode conversion of p-modes. The horizontal axis is the fractional contribution of each model component to the Doppler-velocity power between 3 and 30\mHz. It is derived only from the Bayesian fit to the observed power spectrum (shown in panel a). The contribution from the peak around 6--8\mHz\ is not shown but makes up less than 10\%. Panel c) displays a comparison of observed energy density per logarithmic frequency to the theoretical coronal energy density from photospheric driven Alfvénic fluctuations. The light purple shaded region is the expected energy density estimated by combining a theoretical transmission profile with the observed power spectrum from photospheric bright point motions. The shaded region covers transmission profiles associated with photospheric field strengths from 800--1500\,G.}
\label{fig:decomp}
\end{figure}

Fitting a physics-based model to the Cryo-NIRSP velocity power spectra \cite{Morton2025ApJL} (Methods) separates three components (Figure~\ref{fig:decomp}a). Convectively driven waves dominate, providing $\sim$60\% of the power in 3--30\mHz\ (Figure~\ref{fig:decomp}b). A component near 4\mHz, which may arise from mode conversion of p-modes \cite{Cally2008SoPh}, contributes nearly 30\% in this region. A component occurring between 6--8\mHz\ of unknown origin \cite{Morton2025ApJL} contributes less than 10\%. Both additional timescales have been identified independently in PSP magnetic-field fluctuations (see Figure~\ref{fig:psp}), the 6--8\mHz\ signal regularly \cite{Huang2024ApJL} and the 4\mHz\ signal less often \cite{Huang2026ApJL}. The spectral components resolved by Cryo-NIRSP in the low corona therefore also appear to be present near the Alfvén surface. None of the Cryo-NIRSP targets analysed here lies in unambiguously open field (Supplementary Information). This does not affect the decomposition as the shape of the transmitted spectrum is set by reflection and damping below the corona \cite{Morton2025ApJL,Soler2017ApJ,Soler2019ApJ}, and the Cryo-NIRSP energy densities agree with the open-field UCoMP Doppler power spectrum measured in the same line (compare Figure~\ref{fig:compare} to Figure~\ref{fig:decomp}c; Table~\ref{tab:stats}).

Energy-flux estimates provide an additional line of evidence linking inner-coronal Alfvénic fluctuations to those observed in the near-Sun wind. Assuming the solar wind is primarily accelerated by Alfvénic waves, an upper limit of the wave flux is estimated to be around 50\Wm\ in the corona (see Methods). This value is smaller than that estimated for the coronal Alfvénic waves ($\sim$100\Wm\ \cite{Morton2025ApJ985}) but is consistent with the solar wind waves being remnants of those observed in the inner corona. The difference between the two energy fluxes may indicate that some of the fluctuations observed in the corona have been dissipated to heat the solar wind plasma. Wave heating at these intermediate heights is expected to occur through MHD turbulence \cite{Cranmer2007ApJS}. The result is also in agreement with recent observations showing that a substantial amount of the kinetic energy of the solar wind is from Alfvén waves \cite{Rivera2024Science}.

As mentioned, interchange reconnection has often been suggested as a driver of Alfvénic waves in the corona and solar wind \cite{Bale2023Nature,Chitta2023Science,Raouafi2023ApJ}. However, at the reported rates \cite{Chitta2025AA}, coronal-hole jets perturb only $\sim$1\% of the open flux in 30\,min, so a given field line is disturbed for $\lesssim 0.1\%$ of the time. This rate is clearly far smaller than required to maintain a near ubiquitous and omnipresent wave field. Although, the sources of the solar wind energy and mass flux must be multicomponent. Interchange reconnection may play a substantial role in contributing to the observed intermittency in the wind streams and appears required to explain the observed composition in certain streams \cite{Bale2023Nature}. The associated jets could supply the mass flux \cite{Chen2025AA}. The convective origin of the Alfvénic fluctuations is also supported by recent detailed radiative MHD simulations, which, in favourable conditions, suggest only 30--50\% of the magnetic energy flux comes from reconnection with the majority coming from below the corona \cite{Chen2025AA,Iijima2023ApJL}.

In summary, Alfvénic fluctuations in open coronal fields carry the spectrum predicted for waves driven by surface convection and transmitted through the lower solar atmosphere. This spectrum is present in the cores of polar coronal holes throughout the solar cycle, and in both weakly and rapidly expanding open fields, the likely sources of the fast and slow Alfvénic wind. Its shape is therefore imprinted below the corona and is insensitive to the coronal expansion that determines the wind speed. Averaged over 270 Alfvénic intervals below the Alfvén surface, the young solar wind carries a similar spectrum. This shows that the fossil of convective driving survives to at least the Alfvén surface, as predicted by non-WKB transport theory \cite{Cranmer2005ApJS}, and the wave energy flux in the wind is consistent with it being the remnant of the coronal flux after partial dissipation. High-resolution Cryo-NIRSP spectroscopy further shows that the wave spectrum has several components, with contributions near 4\mHz\ and 6--8\mHz\ alongside the dominant convective component; both timescales also appear in the wind. These correspondences are statistical, between populations rather than individual connected field lines, and coordinated remote-sensing and in-situ observations of a common source will provide the direct test. Taken together with previous studies \cite{Bale2023Nature,Chitta2023Science,Raouafi2023ApJ}, the results indicate that convectively driven Alfvénic waves provide a basal, steady energy supply common to the fast and slow wind, on which interchange reconnection adds intermittent energy and mass.

\begin{table}[t]
\centering
\caption{\textbf{Mean energy density statistics} for each data set, estimated between frequencies of 0.3--15\mHz.}
\label{tab:stats}
\begin{tabular}{lccc}
\toprule
 & Peak frequency & Median frequency & 20--80\% energy range \\
  & (mHz) & (mHz) & (mHz) \\
\midrule
SDO/AIA PCH                        & 5.0 & 4.3 & 2.2--8.2 \\
UCoMP PCH                          & 8.0 & 4.3 & 2.1--8.9 \\
UCoMP ECH                          & 4.5 & 3.6 & 1.8--8.0 \\
PSP average                        & 5.6 & 2.7 & 1.0--7.5 \\
Cryo-NIRSP 30-10-23                & 4.0 & 4.1 & 2.2--7.7 \\
Cryo-NIRSP 23-03-25                & 4.5 & 3.9 & 2.0--7.5 \\
Transmission $\times$ driver (800\,G)  & 5.1 & 3.9 & 2.1--7.2 \\
Transmission $\times$ driver (1500\,G) & 8.2 & 5.9 & 3.4--10.2 \\
\bottomrule
\end{tabular}
\end{table}
\FloatBarrier

\section*{Methods}

\subsection*{SDO/AIA observations and preparation}
This study uses data from the Solar Dynamics Observatory \cite{Pesnell2012SoPh} Atmospheric Imaging Assembly \cite{Lemen2012SoPh} obtained over eight different years during different stages of the Sun's solar cycle. The obtained data are from the Joint Science Operations Centre and are processed to Level 1.5 (i.e., rotation corrected and scaled). Full details of the observations and dates used for the polar coronal hole wave analysis (and additional processing steps) are given in \cite{Morton2025ApJ985}.

We also use data from SDO/AIA on the 2 and 5 October 2021 to support analysis of the UCoMP data. These are Level 1.5 data but are not subject to any analysis.

\subsection*{Alfvénic waves in SDO/AIA}
The Alfvénic waves in the SDO/AIA images are measured by following the fine-scale coronal structure in time-distance diagrams using the NUWT code \cite{Weberg2018ApJ}. This provides both frequency and displacement amplitude estimates for the waves, which we use to convert to velocity amplitude. A detailed breakdown of wave properties across the coronal holes is provided in \cite{Morton2025ApJ985}. To estimate the PSD, we follow the method outlined in \cite{Morton2019NatAstron}. Essentially, the time-averaged power (equivalent to the power of the discrete Fourier transform or global wavelet) for an individual fluctuation can be estimated via
\begin{equation}
P_{i}(f_{i})=\frac{\delta(f)\,T_{i}}{N\,\dd t}\,v_{i}(f_{i})^{2}
\label{eq:aia_power}
\end{equation}
where the subscript $i$ represents an individual measurement with amplitude, $v_{i}$, and frequency, $f_{i}$. Also $T_{i}$ is the duration of each measured signal, $N\,\dd t$ is total length of the observation periods and $\delta(f)$ is the frequency distribution of the waves. Each individual measurement of $P_{i}$ is then used to estimate the power spectrum. The power spectrum can then be converted to the PSD by a constant scaling factor. Some modifications are made to the methodology in \cite{Morton2019NatAstron} to improve the estimate of the power spectrum and associated uncertainties. Firstly, the histogram method for estimating $\delta(f)$ is replaced by Kernel Density Estimation (KDE), where the uncertainty and bias of the estimate are calculated using bootstrapping and the smoothed bootstrap, respectively \cite{Narsky2013Book}. The best bandwidth for the KDE is selected via cross-validation. The power spectrum estimation requires the use of bivariate regression. For this we utilise non-parametric Nadaraya--Watson kernel regression using a radial basis function kernel, with the kernel bandwidth selected by cross-validation. Bootstrapping is used to determine the confidence intervals. The bootstrap process involves: i) calculating the bias-corrected KDE for $\delta(f)$ for each bootstrap sample using the fixed bandwidth and bias found from the full data; ii) re-estimating the individual $P_{i}$ for each bootstrap sample; iii) estimating the bivariate regression curve for each bootstrap sample using the best kernel bandwidth estimated from the full data set. The pointwise confidence intervals are then calculated from the bootstrap regression curves via a percentile method.

\subsection*{Estimating the power spectral density}
Multiple methods are used to examine the power spectral density, $P(f)$, (as a function of frequency, $f$) of the fluctuations across the different data sets. The use of the Discrete Fourier Transform is a common way to calculate the power spectral density (PSD), where the one-sided PSD at frequency $f_{k}$ is given by
\begin{equation}
P(f_{k})=\frac{2}{N f_{s}}\,\lvert X_{k}\rvert^{2}
\label{eq:dft}
\end{equation}
where $N$ is the length of the time-series, $f_{s}$ is the sampling frequency and $X_{k}$ is the $k^{\mathrm{th}}$ Fourier coefficient.

An alternative method is to use Wavelets as the basis functions, instead of cosines as in the case of Fourier analysis. This provides a time frequency representation of the signal, where the wavelet power spectrum, $S(t,f)$, is given by
\begin{equation}
S(t,f) = \lvert W(t,f)\rvert^{2}
\label{eq:wavelet}
\end{equation}
where $W(t,f)$ is the wavelet coefficient at time $t$ and frequency. The so-called global wavelet spectrum is directly comparable to the Fourier power spectral density, and is given by
\begin{equation}
P_{\mathrm{WL}}(f)=\frac{\delta t}{N}\sum_{t=0}^{n-1}\lvert W(t,f)\rvert^{2}
\label{eq:globalwl}
\end{equation}
where $\delta t$ is the sampling interval.

The power per decade of frequency, $P(f)f$, is also calculated for the different time-series. The motivation for calculating this quantity comes from the consideration that the energy stored in moving window of size $\dd f$ is given by
\begin{equation}
\int P(f)\,\dd f=\int P(f)\,f\,\dd\ln f.
\label{eq:edens}
\end{equation}
The second representation (right hand side) shows that the power spectrum multiplied by frequency is proportional to the energy density of frequency $f$ on a logarithmic scale. Furthermore, it can be inferred that the area under the power spectral density in a linear-log plot represents the relative amount of fluctuation energy for a given frequency range. Further discussion can be found in \cite{Cranmer2005ApJS,Liu2007JAOT,Huang2024ApJL}.

\subsection*{Potential field extrapolations}
To support the interpretation of the data and for selecting regions of interest, we undertake potential field modelling of the coronal magnetic field. To do this, we choose to use the ADAPT-GONG \cite{Harvey1996Science} magnetograms (available at \url{https://gong.nso.edu/adapt/maps/gong}) on the 5 October 2021, which have 12 independent realisations of the surface flux transport model used to estimate the evolution of the magnetic field in a Carrington map. The Carrington map magnetograms are used as the lower boundary condition in the \textit{pfsspy} potential field extrapolation tool \cite{Stansby2020JOSS,SunPy2020ApJ}. We initially run the potential field extrapolation using a grid with $1^{\circ}$ spacing in both longitude and latitude for field line seed points. The extrapolations across all the ADAPT maps shows the presence of equatorial open field at the limb (and polar open field) on the 5 October 2021. We then select a $20^{\circ}\times20^{\circ}$ grid around the equatorial open field regions and use a $0.16^{\circ}$ (longitude) and $0.32^{\circ}$ (latitude) spacing for seed points. For the polar field, we choose a much sparser sampling and take points at every $3^{\circ}$ longitude across the Carrington map. The source surface is set at $2.5\,\Rsun$ for the examples shown. We display the results of these extrapolations in Extended Data Figures~\ref{ed:ucomp}a and \ref{ed:pfss}, overplotted on SDO/AIA data. In addition, we calculate the magnetic field expansion factor between the photosphere and source surface, given by
\begin{equation}
f_{ss}=\left(\frac{\Rsun}{R_{ss}}\right)^{2}\frac{B(\Rsun)}{B(R_{ss})}
\label{eq:expansion}
\end{equation}
where $\Rsun$ is the solar radius, $R_{ss}$ is the height of the source surface, and $B(r)$ is the value of the magnetic field at those locations. The expansion factors for the equatorial and polar coronal holes are shown in Figure~\ref{fig:overview}c. We note that using different values of the source surface, e.g., $2.5\,\Rsun$, does not substantially change the distributions of the expansion factor (or the potential field).

\subsection*{UCoMP observations and preparation}
The UCoMP observations for the 5 October 2021 are obtained from the Mauna Loa Solar Observatory web portal (with data doi: \href{https://doi.org/10.26024/G8P7-WY42}{10.26024/G8P7-WY42}). The data have been processed by the instrument team to their Level 2 product, which is science ready. The UCoMP data set consists of measurements of the full Stokes vectors (I, Q, U, V) at a 2.93\arcsec\ spatial sampling. A series of measurements were taken as part of the UCoMP waves programme, which aims to have a high cadence, regularly sampled data set. However, the level 2 has various gaps indicating missing data, with a 76\% duty cycle. We fill these gaps and provide a regularly sampled time-series using cubic-spline interpolation, which has a final cadence of 30.5\,s. An example of an enhanced Stokes I intensity image is shown in Extended Data Figure~\ref{ed:ucomp}b. The image has been enhanced with an unsharp mask filter to better show the variations in brightness.

We also use the Stokes Q and U vectors from the synoptic average measurements. The ratio of these two polarisation components gives the plane of sky component of the magnetic field
\begin{equation}
\phi=\frac{1}{2}\arctan\left(\frac{U}{Q}\right).
\label{eq:azimuth}
\end{equation}
We convert this value into an angle with respect to the local radial direction. For the 5 October 2021 the map of radial azimuth is shown in Extended Data Figure~\ref{ed:ucomp}c. We also calculate the fractional linear polarization $\sqrt{Q^{2}+U^{2}}/I$ and this is shown in Extended Data Figure~\ref{ed:ucomp}d.

\subsection*{UCoMP analysis}
Two regions of the UCoMP data are selected for further analysis. These correspond to the rapidly expanding equatorial coronal hole identified in the potential field extrapolations, and the polar coronal hole. The equatorial hole is visible in the UCoMP enhanced intensity image (Extended Data Figure~\ref{ed:ucomp}b), showing brightness enhancements oriented near radially, which is confirmed by the measurements of the radial azimuthal magnetic field angle in Extended Data Figure~\ref{ed:ucomp}c. The equatorial coronal hole also has a distinct signal in the fractional linear polarisation. As noted in the main text, \FeXIII\ emission is weak in the cores of polar coronal holes as they are relatively cool ($\sim$0.7\,MK) compared to the peak formation temperature of the line (1.6\,MK). Hence, it is challenging to observe large polar coronal holes with UCoMP. However, there is a small region at the base of the polar coronal hole which has measurable signal.

The analysis for the Alfvénic waves only uses the Doppler velocities. Each time-series in the boxed fields of view is subject to a Tukey window (with $\alpha=0.1$), then a wavelet transform. We then calculate the global wavelet power, summing only over the power outside of the cone of influence. These are the power spectral densities shown in Figure~\ref{fig:overview}b.

\subsection*{PSP observations and preparation}
Here we use the magnetic field data from the fluxgate magnetometer in the FIELDS instrument suite \cite{Bale2016SSRv} and plasma moments from SPAN-i and SPAN-a in the SWEAP instrument suite \cite{Kasper2016SSRv}. The radial component of the first-order proton (partial) moment is used as an estimate for the solar wind speed. The Quasi-Thermal Noise (QTN)-derived electron number density, $N_{e}$, is taken as a proxy for particle number density (composed of alpha particles, $N_{\alpha}$, and protons, $N_{p}$). Assuming charge neutrality ($N_{e}=2N_{\alpha}+N_{p}$), the mass density is calculated as: $\rho=N_{e}m_{p}$.

Data are selected from Encounters 10--24 and are taken from 3 days of each perihelion encounter. We further require that the heliocentric distance of PSP is below $15\,\Rsun$ and is below the Alfvén surface (measured by the Alfvén Mach number $M_{A}=V_{R}/V_{A}$, where $V_{R}$ is the radial velocity and $V_{A}$ is the Alfvén speed). Intervals of the data were selected that have an absolute cross helicity $\vert\sigma_{c}\vert>0.8$, hence are Alfvénic in nature. In total, we analyse 270 individual 2-hour windows. Extended Data Figure~\ref{ed:psp} shows an example fast wind stream in full detail.

The cross helicity $\sigma_{c}$ is used to estimate the Alfvénicity of the solar wind timeseries. $\sigma_{c}$ is calculated in two forms: $\sigma_{c}(t)$, which shows the temporal correlation between the magnetic field, $\bm{B}$, and velocity, $\bm{V}$; and $\sigma_{c}(f)$, which express the frequency-dependent correlation. First, the magnetic field is converted into velocity units:
\begin{equation}
\bm{V}_{A} = \frac{\bm{B}}{\sqrt{\mu_{0}\rho}}.
\label{eq:va}
\end{equation}
The fluctuating components are obtained by subtracting the 10-minute rolling average:
\begin{align}
\delta V_{A,\{R,T,N\}} &= V_{A,\{R,T,N\}}-\langle V_{A,\{R,T,N\}}\rangle, \label{eq:dva}\\
\delta V_{\{R,T,N\}} &= V_{\{R,T,N\}}-\langle V_{\{R,T,N\}}\rangle. \label{eq:dv}
\end{align}
And the cross-helicity is defined as
\begin{equation}
\sigma_{c}(t)=\frac{2\,\delta\bm{V}\cdot\delta\bm{V}_{A}}{\lVert\delta\bm{V}\rVert^{2}+\lVert\delta\bm{V}_{A}\rVert^{2}}.
\label{eq:sigc_t}
\end{equation}
To compute $\sigma_{c}(f)$ we use the Fourier transform of the velocity and magnetic field instead, i.e.
\begin{equation}
\sigma_{c}(f)=\frac{2\,\mathcal{F}(\delta\bm{V})\cdot\overline{\mathcal{F}(\delta\bm{V}_{A})}}{\lVert\mathcal{F}(\delta\bm{V})\rVert^{2}+\lVert\mathcal{F}(\delta\bm{V}_{A})\rVert^{2}}.
\label{eq:sigc_f}
\end{equation}

\subsection*{Parker Solar Probe PSD}
The PSD for Parker Solar Probe is calculated from the trace of the magnetic field fluctuations, which are resampled to 0.5\,s. Denoting the Fourier transform of the magnetic field fluctuations as $\hat{B}_{\{r,t,n\}}(f)$, then the Fourier trace magnetic PSD is computed as:
\begin{equation}
P(f)=\frac{2}{N f_{s}}\left(\lvert\hat{B}_{r}(f)\rvert^{2}+\lvert\hat{B}_{t}(f)\rvert^{2}+\lvert\hat{B}_{n}(f)\rvert^{2}\right).
\label{eq:psp_fft}
\end{equation}
For the wavelet estimate of the power spectral density, the magnetic field fluctuations are first transformed into individual wavelet spectrum, $S_{\{r,t,n\}}(t,f)$. Again, the Morlet wavelet ($f_{0}=6$) is used as the mother wavelet. The trace wavelet PSD is then calculated as:
\begin{equation}
P_{\mathrm{WL}}(f)=S_{r}(t,f)+S_{t}(t,f)+S_{n}(t,f)
\label{eq:psp_wl}
\end{equation}
This is converted to the global PSD following Eq.~\ref{eq:globalwl}. Examples of the Fourier and wavelet trace magnetic PSDs are given in Figure~\ref{fig:psp}c and f.

\subsection*{Relating PSP frequencies to UCoMP \& SDO}
It is known that, for a wave forced at frequency $\omega$, the temporal frequency remains unchanged as it travels. In the case of the corona, as the solar wind speed, $U(r)$, increases with distance from the surface, it effectively stretches out the parallel wavelength associated with a propagating wave. Hence, for non-dispersive Alfvén waves, the dispersion equation is $\omega=(U+V_{A})k$ and the wavenumber is altered as the wave propagates outwards.

As PSP makes its Encounters with the Sun, it has components of motion both parallel and perpendicular to the local radial direction with respect to the Sun ($V_{R,\mathrm{PSP}}$, $V_{\theta,\mathrm{PSP}}$). There are two aspects to consider. The first is related to the radial velocity. It has been demonstrated \cite{Huang2024ApJL} that if PSP's radial velocity is negligible compared to the propagation speed of the fluctuations, then there is practically no Doppler shift of the observations. For example, when the radial velocity of PSP is $V_{R,\mathrm{PSP}}\sim100\kms$ and the fluctuations have a phase speed of $v_{A}+U\sim1000\kms$ (Extended Data Figure~\ref{ed:psp}f) then the Doppler shift is on the order of a factor $\omega_{\mathrm{psp}}=1.1\,\omega_{0}$ (where $\omega_{0}$ is the driven frequency at the Sun), leading to a 10\% difference. In most cases, the radial velocity of PSP is small compared to the fluctuation propagation speed, leading to a median $\sim$6\% change in frequencies. When PSP is close to perihelion, PSP is usually moving tangential to the Sun (practically perpendicular to the magnetic fields) and $V_{R,\mathrm{PSP}}\approx0$. PSP is practically in the Sun’s rest frame, hence, there is no Doppler shift of the observations.

The other consideration arises from the perpendicular motions of PSP, which can turn spatial structures into temporal signals \cite{Huang2024ApJL}. The perpendicular contamination frequency is given by
\begin{equation*}
f_{c}=\frac{V_{\theta,\mathrm{PSP}}}{3\lambda\,(R/\Rsun)}
\end{equation*}
where $\lambda$ is the spatial scale of the structures and assuming a super radial expansion factor of 3. During the analysed streams, $V_{\theta,\mathrm{PSP}}=[120,180]$\kms\ and $R=[10,15]\,\Rsun$. The PSP energy density per logarithmic frequency shows a peak at $\sim$6\mHz\ and 60\% of the energy is between (1--30\mHz\ when using full frequency range). To generate a peak at 6\mHz, this would specify a preferred spatial scale of the magnetic field consistent with 20,000--30,000\,km at 10--15\,\Rsun\ and $\lambda\sim$600\,km in the inner corona. While 600\,km is broadly consistent with photospheric granulation, due to the significant expansion between photosphere and corona (magnetic expansion factors between the photosphere and corona are around $\sim$50) this would correspond to scales of $\sim$100\,km at the photosphere, which is the average size of individual magnetic flux tubes in the intergranular lanes. However, it is expected that the individual flux tubes merge in the chromosphere and the magnetic field is relatively homogeneous over supergranule scales in the corona \cite{Cranmer2005ApJS}. Remote magnetic field measurements of the corona are still challenging and do not yet have the capability to detect variations over small spatial scales. The fine structure observed in remote imaging of the inner corona is thought to be composed of density enhancements, with spatial power spectra for the intensity perpendicular to magnetic field consistent with a $k^{-1}$ scaling in coronal holes \cite{Morton2023ApJ}. This scaling of intensity features at 10\,\Rsun\ also shows the same scaling down to the resolution limit of 20,000\,km \cite{DeForest2018ApJ}. Hence, there is little evidence for the magnetic field being structured on such scales as to cause a contamination.

\subsection*{Energy density statistics}
We compute the energy density statistics by selecting a common frequency interval across all the observational data sets. The UCoMP measurements have a Nyquist frequency of around 16\mHz\ and the UCoMP and Cryo-NIRSP data sets are around an hour long, giving a minimum frequency of 0.2\mHz. Hence, we choose an interval of 0.3--15\mHz\ to calculate the statistics.

Beyond 20\mHz, the Cryo-NIRSP Doppler velocity power becomes increasingly dominated by white noise (Figure~\ref{fig:decomp}a), which would bias results. The PSP data are analysed in 2 hour windows with a reduced sampling of 0.5\,s. This means that magnetic field fluctuations are well above the noise floor for the FIELDS instrument. Calculating the energy density statistics over this larger window shows the median value and upper edge of the 20--80\% frequency band are sensitive to the window of evaluation. As noted in the main text, Figure~\ref{fig:compare} shows that PSP energy density is broader than the coronal profiles.

\subsection*{DKIST observations and preparation}
The DKIST observations that are discussed in Figures~\ref{fig:compare} and \ref{fig:decomp} were taken by the Cryo-NIRSP \cite{Fehlmann2023SoPh} instrument, which can observe the corona in the infrared using the \FeXIII\ 1074\,nm emission line. The data were obtained from the DKIST Data Centre (\url{https://dkist.data.nso.edu/}), where they have been processed to Level 1 by the DKIST data pipeline, which includes various corrections required to make the data of science quality (as detailed in the data product documentation). Our analysis uses four spectroscopic datasets (2 observing days with two different observing modes), with additional details given in the Supplementary Information. To extract the coronal line parameters such as line amplitude and Doppler velocities, the local infrared spectrum in the region of 1074\,nm must be modelled accurately. This includes incorporating scattered photospheric light and telluric absorption lines into the spectral model. Full details of how this is performed are given in \cite{Schad2023ApJ,Schad2024ApJ}. The coronal line within the spectrum is modelled with a single Gaussian profile. Examples are given in the Supplementary Information.

The cadence for the sit and stare data is approximately one second. The spatial sampling along the Cryo-NIRSP slit is 0.12\,arcseconds. However, estimates of seeing quality based on the coherency of fluctuations \cite{Morton2025ApJ982} suggest that the spatial resolution is on the order of 0.4\,arcseconds due to terrestrial seeing conditions.

\subsection*{Cryo-NIRSP Doppler velocity PSD and modelling}
The wavelet representation of the Doppler velocity PSD, namely the global wavelet PSD, is estimated for the Cryo-NIRSP data to derive the energy density per logarithmic frequency, $P_{\mathrm{WL}}(f)f$. A Morlet wavelet ($f_{0}=6$) is used as the mother wavelet. This is done for multiple time-series independently and then an average energy density is estimated, and is shown in Figure~\ref{fig:decomp}c. For each data set, we limit this averaging to regions in which we believe are dominated by emission associated with open field regions. However, as discussed in the Supplementary Information, we cannot guarantee there is no contamination from closed field emission. For the March 2025 data, pixels located at $\gtrsim$168\arcsec\ along the slit are used for the analysis in Figure~\ref{fig:decomp}. This corresponds to a wine-glass shaped geometry defined by increased non-thermal line widths compared to the closed loops on the left-hand side of the image (Extended Data Figure~\ref{ed:cryo}). Potential field extrapolations also support the potential presence of open field in this region (Supplementary Information).

Figure~\ref{fig:decomp}a shows an example of the decomposition of the observed Doppler velocity power spectrum for the data taken on March 2025. For fitting a model to the Cryo-NIRSP data, the power spectral density for the Doppler velocity fluctuations is calculated by the discrete Fourier transform. The average Fourier power spectrum is then estimated for the same regions used for the energy density analysis. The uncertainties on the power spectrum are calculated following \cite{Morton2019NatAstron} which includes a correction for the loss of independence of neighbouring time series due to the terrestrial seeing.

The modelling of the coronal power spectrum follows that outlined in \cite{Morton2025ApJL}. In brief, the model contains components to describe the power law, the data noise, contribution from Alfvénic waves driven by the convective motions at the surface and two Gaussian functions to describe the additional contributions. The term in the model corresponding to the power spectrum from convective motions is a combination of a theoretical profile describing the wave energy flux transmission into the corona from Alfvénic waves excited at the photosphere \cite{Soler2019ApJ} and a measured power spectrum of magnetic bright point motions measured in the photosphere \cite{Chitta2012ApJ}. The transmission profile includes the effects of wave reflection from the transition region and wave damping in the chromosphere from partial ionization and phase mixing. The transmission profile is parameterised as a function of the photospheric magnetic field, which is left as a free parameter in the model. Different values of the photospheric magnetic field can influence both wave reflection and the wave damping, which the parametrised transmission profile captures (assuming the model of Soler et al.\ \cite{Soler2019ApJ} is an accurate description). The model is fit to the data using a Bayesian methodology and we use a Hamiltonian Monte Carlo method \cite{AbrilPla2023PeerJCS} to sample the posterior. The results of the modelling are displayed in Figure~\ref{fig:decomp}a for March 2025. The posterior gives $B=1158_{-541}^{+606}$\,G (mean and 95\% posterior highest density interval).

We then sample models from the joint posterior distribution to estimate the fractional contribution to the velocity power from the different model components, i.e., the transmission-driver term and the two Gaussians (results shown in Figure~\ref{fig:decomp}b). Each posterior sample represents a different combination of parameters for each of these components. The power associated with each component is individually summed and divided by the total power across the 3 components, giving the fractional contribution to the Doppler velocity power per component. Doing this for each posterior sample provides the posterior distributions for the fractional contribution to velocity power from each component.

\subsection*{Solar wind energy budget}
In a time-steady, radially expanding solar wind stream, the energy flux is conserved on a given cross section \cite{Belcher1971ApJ} (in c.g.s.\ unit):
\begin{equation}
F=4\pi r^{2}\left\{V\left[\frac{1}{2}\rho V^{2}+\frac{\alpha}{\alpha-1}p-\frac{GM_{\odot}\rho}{r}\right]+\frac{1}{2}\left(\frac{1}{2}\rho\,\delta V^{2}\right)V+\frac{1}{2}\frac{\delta B^{2}}{4\pi}\,(V+V_{A})\right\}
\label{eq:flux}
\end{equation}
where $V$ is the proton bulk speed, $V_{A}$ is the local Alfvén speed. The pressure and density are assumed to satisfy the polytropic relation:
\begin{equation}
p=p_{0}\left(\frac{\rho}{\rho_{0}}\right)^{\alpha}
\label{eq:poly}
\end{equation}
and $\alpha$ is the polytropic index. In the case of monoatomic ideal gas, $\alpha=5/3$, but in solar wind plasma, this value often deviates from $5/3$ substantially \cite{Shi2022PhPl}. In the ideal Parker solar wind model \cite{Parker1958ApJ}, the solar wind is primarily accelerated by the gas pressure gradient, which is the second term in Eq.~\ref{eq:flux} (also known as the enthalpy flux):
\begin{equation}
F_{H}=\frac{\alpha}{\alpha-1}\,pV
\label{eq:enthalpy}
\end{equation}
However, the Parker model fails to reproduce the anti-correlation between the solar wind speed and density in the fast solar wind that originates from coronal holes, which is often filled with large amplitude Alfvén waves. Thus, it is believed that the Alfvén wave energy flux provides a significant amount of energy, if not a dominant contribution, to accelerate the plasma in the fast solar wind.

In the fast solar wind, the fluctuations are primarily Alfvénic, which show a near-perfect correlation between $\delta V$ and $\delta B$. Therefore, the last two terms can be combined into a single Alfvén wave energy flux ($F_{w}$):
\begin{equation}
F_{w}=\frac{1}{2}\frac{\delta B^{2}}{4\pi}\left(\frac{3}{2}V+V_{A}\right)
\label{eq:fw}
\end{equation}
As the solar wind accelerates from the coronal base outwards into the heliosphere, the enthalpy and wave energy flux decrease to increase the kinetic energy flux ($F_{K}$) and to overcome the gravitational well ($F_{G}$):
\begin{equation}
F_{K}=\left(\frac{1}{2}\rho V^{2}\right)V,\qquad F_{G}=\left(-\frac{GM_{\odot}\rho}{r}\right)V.
\label{eq:fkfg}
\end{equation}
The differences of $F_{K}$ and $F_{G}$ between the spacecraft and the coronal base reveal the amount of energy in the solar wind that is gained from $F_{W}$ and $F_{H}$. Therefore, the following value can be regarded as the upper limit of $F_{W}$ at the coronal base (in the limit that the contribution of $F_{H}$ is minimal):
\begin{equation}
F_{W}\big\rvert_{R=\Rsun}\leq F_{W}\big\rvert_{R}+(F_{K}+F_{G})\big\rvert_{R}-(F_{K}+F_{G})\big\rvert_{R=\Rsun}.
\label{eq:bound}
\end{equation}
Here, the notation $\rvert_{R}$ indicates the flux is evaluated at a distance $R$ from the solar surface. It should be noted the $F_{K}$ also has a contribution from the kinetic energy flux carried by the alpha particles, which contains a comparable amount of kinetic energy flux to that carried by protons due to the significant differential streaming \cite{Rivera2024Science,Wang2025ApJL}. In this work, based on the fast wind event in Extended Data Figure~\ref{ed:psp}, the upper limit of $F_{w}$ is estimated to be about 50\Wm.

\subsection*{Energy flux of Alfvénic waves from remote sensing}
While we do not make any estimates of the coronal Alfvénic energy flux here (estimates are taken from \cite{Morton2025ApJ985}, which use the same SDO/AIA data), it is worth discussing briefly the associated issues. With Doppler velocity data from coronal emission lines, the line of sight integration of plasma emission through the optically thin corona leads to a reduction in measured fluctuation amplitudes due to summation over incoherent sources -- this is a well appreciated effect (see discussions in e.g., \cite{Morton2026NatAstron}). Imaging observations do not suffer as obviously from this effect. The imaging data appear to show fine-scale, field-aligned enhancements in brightness. This is generally considered to represent enhanced density associated with individual, discrete coronal structures. Although the integration through the optically thin coronal plasma may lead to the appearance of discrete structures when none exist \cite{Malanushenko2022ApJ}. Even then, the plasma elements contributing to the increase in brightness are still relatively localised ($<$10\,Mm in the simulations of \cite{Malanushenko2022ApJ}). Hence, the observed transverse displacements observed in SDO/AIA data are expected to be related to localised plasma enhancements in the corona, and do not suffer considerable reduction in amplitude. Conservatively, we can expect the measured Alfvénic energy fluxes in the corona to be lower bounds.

\subsection*{Can picoflares be the main driver for the observed Alfvénic fluctuations?}
We suggest there are several reasons that picoflares (or other coronal jets) cannot be the main source of Alfvénic fluctuations.

One of the main challenges is related to the energy flux associated with the jets. Estimates for the Alfvénic waves energy flux is at least 100\Wm\ ($10^{5}$\,\ergcgs). It was reported \cite{Chitta2025AA} that picoflare jets numbered around 120 in a 9000\,Mm$^{2}$ region over 30 minutes, with jets widths of 1\,Mm. Hence, in a region with this area, the power required is $9\times10^{24}$\,erg\,s$^{-1}$. Over 30 minutes, the per event jet energy is then $1.4\times10^{26}$\,erg. Estimated kinetic energies for the jets are $10^{21}$--$10^{24}$\,erg, which leaves a short fall of a factor of $10^{2}$--$10^{5}$.
Being conservative and assuming only 10\% of the released energy is converted to kinetic energy, this still leaves a short fall of at least 10 times. Also, only a fraction of this energy would end up as Alfvénic waves. Further, simulations indicate Poynting flux values of $10^{6}$\,\ergcgs\ per event \cite{Chen2025AA}, which given the jets have a spatial filling factor of around 0.01 \cite{Chitta2025AA}, indicates contributions on the order of $10^{4}$\,\ergcgs. Although, as mentioned, in certain simulated situations interchange reconnection provides 30--50\% of the coronal Poynting flux \cite{Chen2025AA,Iijima2023ApJL}.

We can also consider what the coverage (or duty cycle) of the wave excitation may be from the jets. The coronal observations show Alfvénic fluctuations are essentially present everywhere and all the time \cite{Tomczyk2007Science,DePontieu2007Science,McIntosh2011Nature,Morton2019NatAstron,Morton2025ApJ985,Morton2026NatAstron}. Interchange reconnection will only excite Alfvén waves along open flux involved in the reconnection. Given the jet numbers from \cite{Chitta2025AA}, the event rate per unit area is $e_{A}\approx7.4\times10^{-6}$\,Mm$^{-2}$\,s$^{-1}$. The duty cycle of the events is given by $d=e_{A}A_{e}T$ where $A_{e}$ ($\approx0.79$\,Mm$^{2}$) is the per event area of open flux associated with the jet and $T$ is the duration of the wave packet it launches. Hence $d=6\times10^{-4}$ assuming the wave packet time-scale corresponds to the EUV lifetime of the jet (100\,s). Even if the wave packets ring out for 10x the visible lifetime of the jet the predicted duty cycle is still far too low. The mean time between hits of a given open field line is around 50 hours. To obtain a reasonable duty cycle ($d>0.3$) there would need to be 50x to 500x more jets than observed in \cite{Chitta2025AA}.

For waves launched by interchange reconnection, the timescale of the disturbance is set by the energy-release time, which for an impulsive event is of order the Alfvén crossing time of the reconnecting region, $\tau\sim L/v_{A}$. Producing power at 3--10\mHz\ (100--330\,s) would require regions with $L\approx$ 40--230\,Mm for $v_{A}\approx$ 400--700\kms\ \cite{Yang2020Science}. It seems rather arbitrary that such a specific sub-population of loops would be associated with reconnection driven Alfvén waves. Moreover, loops of this length are common in the closed corona, but the closed field embedded in coronal holes is dominated by small, low-lying loops \cite{Wiegelmann2004SoPh}. The similar spectral shape is observed in open and closed fields throughout the corona \cite{Morton2019NatAstron}, despite large differences in loop lengths and Alfvén speeds between environments. This is expected if the shape is set below the corona, by photospheric driving filtered through the chromosphere and transition region. And is difficult to reconcile with a timescale set by coronal reconnection, which would depend upon the local ratio of loop length to Alfvén speed.

\section*{Data Availability}
The SDO/AIA data are available at \href{http://jsoc.stanford.edu/}{jsoc.stanford.edu}. The UCoMP data are available at \url{https://www2.hao.ucar.edu/mlso/mlso-data-and-movies}. For Parker Solar Probe the data are available at \url{https://spdf.gsfc.nasa.gov/pub/data/psp/}. The orbital data are available at \url{https://psp-gateway.jhuapl.edu/}. GONG data are available at \url{https://gong.nso.edu/data/}. The DKIST data used for this study are available at: \url{https://nso.edu/dkist/data-center/}. The data have identifiers L1-EXJWQ, L1-FCHTW, L1-LWOTP, and L1-HMQEO.

\section*{Code Availability}
Code used for the remote observation data analysis is available at \url{https://github.com/Richardjmorton/NUWT} for analysis of Alfvénic waves in images and \url{https://github.com/tschad/cryonirsp-notebooks} for processing of Cryo-NIRSP data. The code for fitting the coronal power spectrum with a Bayesian model is available at \url{https://github.com/Richardjmorton/Bayesian_fit_to_power_spectrum}. The code for the in-situ data analysis is available at \url{https://github.com/huangzesen/SWTurbPy/}.

\section*{Acknowledgements}
The research reported herein is based in part on data collected with the Daniel K. Inouye Solar Telescope (DKIST), a facility of the National Solar Observatory (NSO). NSO is managed by the Association of Universities for Research in Astronomy, Inc., and is funded by the National Science Foundation. DKIST is located on land of spiritual and cultural significance to Native Hawaiian people. The use of this important site to further scientific knowledge is done so with appreciation and respect. UCoMP data are courtesy Mauna Loa Solar Observatory, operated by the High Altitude Observatory, National Center for Atmospheric Research (NCAR), and supported by the National Science Foundation. This work utilizes data from the U.S. National Science Foundation (NSF) Global Oscillation Network Group (GONG), built and operated by the NSF National Solar Observatory (NSO) Integrated Synoptic Program (NISP). NSO is managed by the Association of Universities for Research in Astronomy, Inc. (AURA), and funded under a cooperative agreement with the NSF. GONG is partially supported by the National Oceanic and Atmospheric Administration (NOAA). The GONG network of instruments is hosted by the Big Bear Solar Observatory, High Altitude Observatory, Learmonth Solar Observatory, Udaipur Solar Observatory, Instituto de Astrofísica de Canarias, and Cerro Tololo Inter-American Observatory. R.J.M is supported by the UKRI Future Leaders Fellowship (RiPSAW---MR/T019891/1).

\section*{Author Contributions}
RJM and TS designed the DKIST observing proposal and performed the analysis of the observational data. RJM, ZH and MV collated and analysed the data from PSP. All co-authors contributed to the scientific discussion of the results and writing of the manuscript.

\section*{Competing Interests}
The authors declare no competing interests.

\FloatBarrier
\bibliographystyle{unsrtnat}
\bibliography{references}

\clearpage
\section*{Extended Data}

\begin{edfigure}[t]
\figbox[0 0 0 0]{EDFig1}{11cm}{Polar coronal hole (PCH) velocity power spectra from SDO/AIA (\FeIX\ 17.1\,nm), one panel per year (eight years), with individual measurements, KDE-estimated PSD and confidence intervals; vertical dashed lines at 4 and 8\,mHz.}
\caption{\textbf{Polar coronal hole (PCH) velocity power spectrum derived for Alfvénic motions observed in the plane of sky using the Solar Dynamics Observatory} (measurements made in \FeIX\ 17.1\,nm channel). The year in which the measurements were taken is given in the panel titles. The violet dots are associated with the individual measurements of the wave motions. The purple line shows the power spectral density (PSD) estimated using Kernel Density estimation (KDE) and the confidence intervals are the fuchsia shaded regions. The vertical dashed lines indicate frequencies of 4\mHz\ and 6--8\mHz\ found in the DKIST Doppler velocity power spectrum (Figure~\ref{fig:decomp}a) -- corresponding to approximate frequency locations of the additional power enhancements likely corresponding to additional wave sources.}
\label{ed:aia}
\end{edfigure}

\begin{edfigure}[t]
\figbox[0 0 0 0]{EDFig2}{9cm}{Panels (a)--(d): potential-field lines on SDO/AIA 19.3\,nm (a); UCoMP 1074\,nm intensity with analysis regions (b); radial azimuth (c); fractional linear polarisation (d).}
\caption{\textbf{Supporting data for the UCoMP observations.} The estimated magnetic field for the equatorial coronal hole are displayed on an SDO/AIA (19.3\,nm) image (panel a). The regions used for analysis of Alfvénic fluctuations of the polar and equatorial coronal holes are marked on the 1074\,nm line intensity image from UCoMP (panel b). The radial azimuth (panel c) and the fractional linear polarisation (panel d) also indicate the orientation of the magnetic field.}
\label{ed:ucomp}
\end{edfigure}

\begin{edfigure}[t]
\figbox[0 0 0 0]{EDFig3}{8cm}{Panels (a)--(b): potential-field lines of the equatorial coronal hole on SDO/AIA 19.3\,nm, 2 October 2021 (a); polar coronal hole on SDO/AIA 19.3\,nm, 5 October 2021 (b).}
\caption{\textbf{The magnetic field in the coronal holes estimated from the potential field extrapolations.} The equatorial coronal hole magnetic field is overplotted on an SDO/AIA (19.3\,nm) from the 2 October 2021 (panel a), when the hole is on disk. The polar hole magnetic field is overplotted on an SDO/AIA (19.3\,nm) from the 5 October 2021 (panel b), which is the date of the UCoMP observations.}
\label{ed:pfss}
\end{edfigure}

\begin{edfigure}[t]
\figbox[0 0 0 0]{EDFig4}{12cm}{Panels (a)--(i): PSP Encounter 17 (27-09-2023) fast-wind stream: B$_R$; cross-helicity; trace-B PSD (Fourier, wavelet); field components; energy density; V$_R$ and V$_A$; $\sigma_c$ and plasma $\beta$; proton and alpha densities; Alfv\'en Mach number and radial distance.}
\caption{\textbf{Typical fast wind stream example from Parker Solar Probe.} During this encounter (E17 on 27-09-2023), PSP is around 12\,\Rsun\ from the Sun flying through a fast wind stream (panel a). The wind is highly Alfvénic, with a strong cross helicity across all frequencies (panel b). The power spectral density (PSD) of the trace magnetic field fluctuations (panel c) shows the energy-containing ($<10^{-2}$\,Hz) scales and inertial range ($>10^{-2}$\,Hz). Both the Fourier (blue) and global wavelet (orange) estimates are shown. Also displayed are the compensated power spectrum on a log (orange dashed) and linear (grey dashed dot) vertical scale. Various properties of the solar wind are given for context: (d) the individual time-series for the magnetic field components; (e) the wavelet energy density per logarithmic frequency of the trace magnetic field; (f) Radial solar wind $V_{R}$ and Alfvén speed $V_{A}$; (g) Cross-helicity $\sigma_{c}$ (left; blue) and plasma $\beta$ (right; black). (h) Proton density $n_{p}$ (left; black) and alpha particle density (right; orange); (i) The Alfvén Mach number (red) and the radial distance of PSP from the Sun (black).}
\label{ed:psp}
\end{edfigure}

\begin{edfigure}[t]
\figbox[0 0 0 0]{EDFig5}{9cm}{Panels (a)--(c): Cryo-NIRSP \FeXIII\ 1074\,nm raster: peak line amplitude (a); mean-subtracted Doppler shift (b); non-thermal line width (c).}
\caption{Panel a shows the \FeXIII\ 1074\,nm peak line amplitude (units are millionths of disk brightness, $B_{\odot}$) derived from high-resolution raster-scan data of a region of the corona taken with the DKIST/Cryo-NIRSP instrument. Panels b and c show the corresponding mean subtracted Doppler shift and the non-thermal component of the line width, respectively. The non-thermal velocity is defined as the 1/e half-width of the Gaussian fit converted to velocity, after subtraction in quadrature of the instrumental width and the thermal width (at $T=1.6$\,MK). Bright coronal loops are observed on the left-hand side. A potential open field region is present at the right-hand side ($\gtrsim$150\,arcseconds), indicated by the apparent funnel structure present in the non-thermal line widths (panel c). A full discussion of the magnetic geometry is given in the Supplementary Information.}
\label{ed:cryo}
\end{edfigure}
\FloatBarrier

\clearpage
\section*{Supplementary Information}
\begin{center}
\textit{A magneto-convective origin for solar-wind Alfvénic fluctuations}\\[2pt]
R.~J.~Morton, Z.~Huang, T.~Schad, M.~Velli
\end{center}

Here we discuss the results from investigations into the nature of the coronal regions we observed with DKIST/Cryo-NIRSP.

\subsection*{Details of DKIST/Cryo-NIRSP data sets}
In all the data sets used, there were two data acquisition modes run relevant to this study.

The data set from the 23-03-2025 has product identifiers of \href{https://dkist.data.nso.edu/product/L1-EXJWQ}{L1-EXJWQ} and L1-FCHTW. The first is a raster scan of the region in the low corona (L1-EXJWQ), which is shown in Extended Data Figure~\ref{ed:cryo}. The second set of observations was a sit and stare mode (L1-FCHTW), where the mid-point of the Cryo-NIRSP slit was oriented tangentially with respect to the solar limb. Example spectrum and results of fitting the line are shown in Supplementary Figure~\ref{sf:spec}. The Doppler velocities derived from the fit are shown in Supplementary Figure~\ref{sf:doppler}. The Doppler velocities used for the analysis presented in Figure~\ref{fig:decomp} are taken from $\geq$168\arcsec\ along the slit, corresponding to an estimated location of (1065\arcsec, $-$199\arcsec) to (1057\arcsec, $-$261\arcsec) in helioprojective Cartesian coordinates.

The second data set was from the 30-10-2023 and full details are given in Morton et al.\ \cite{Morton2025ApJ982}.

\begin{sifigure}[t]
\figbox[0 0 0 0]{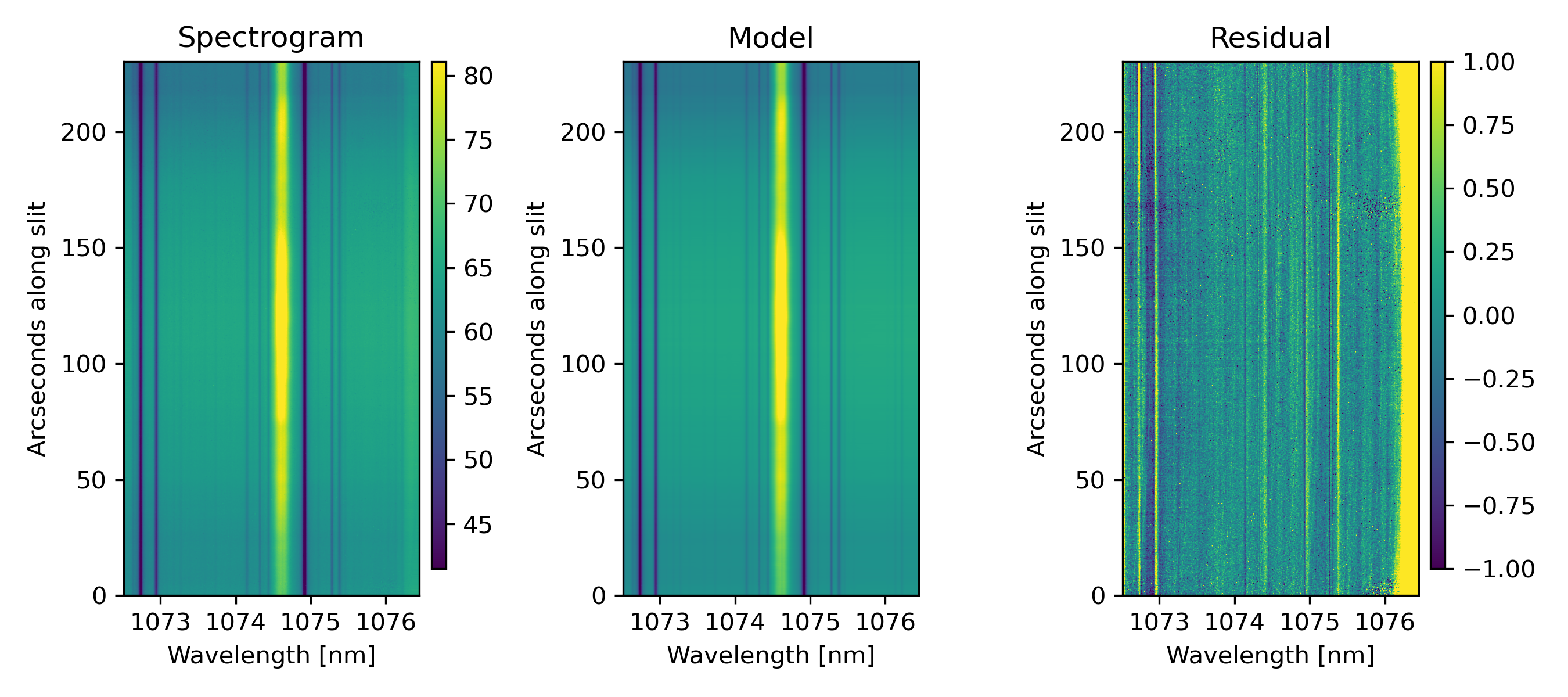}{7cm}{Three panels: observed Cryo-NIRSP coronal spectrum along the slit; model from the fit; residual (units: millionths of disk brightness).}
\caption{Example of the Cryo-NIRSP coronal spectrum along the slit (first panel), the resulting model from the fit to the spectrum (second panel), and the residual between the observed and the model spectra. Units are in millionths of disk brightness. The spectra are from the raster along the pink dashed line shown in Extended Data Figure~\ref{ed:cryo}.}
\label{sf:spec}
\end{sifigure}

\begin{sifigure}[t]
\figbox[0 0 0 0]{SFig2}{7cm}{Left: Doppler-velocity time series along the slit (23-03-2025 sit-and-stare). Right: two example time series at 170$''$ and 200$''$ along the slit.}
\caption{Doppler velocity from the Cryo-NIRSP sit and stare observing sequence on the 23-03-2025. The left panel shows the full Doppler velocity time-series along the slit. The right panel shows two example time-series at locations 170\arcsec\ and 200\arcsec\ along the slit.}
\label{sf:doppler}
\end{sifigure}

\subsection*{Magnetic geometry of regions observed with Cryo-NIRSP}
For each case we use a GONG synoptic magnetogram with a time-stamp within or near to the DKIST observation window. Potential field extrapolations are used to calculate maps of open and closed flux at the photosphere based on whether the field line is open or closed within the defined source surface. Various values of source surface height are investigated as there is now significant evidence \cite{Panasenco2020ApJS,Badman2020ApJS} that a single value of $R_{ss}\approx2.5\,\Rsun$ (the usual default value) in potential field models does not consistently agree with other observational diagnostics (e.g., observed coronal holes in imaging, magnetic field polarity observed in-situ).
Seeds for the extrapolation points are distributed evenly across longitude and latitude at the resolution on the magnetogram.

Synoptic maps of the EUV corona were constructed from Level 1 19.3\,nm images from the SDO/AIA taken at a cadence of 6\,h during a Carrington rotation (CR). Each image was processed to level 1.5. The images were then binned $4\times4$ to 2.4\arcsec\ pixel$^{-1}$. To reduce the centre-to-limb variation in intensity, on-disk pixels were grouped into 50 equal-width bins of $\mu=\cos\theta$, where $\theta$ is the heliocentric angle. Each bin was rescaled so that its median matched that of the disk-centre bin ($\mu>0.98$), with the correction factor interpolated linearly in $\mu$ between bin centres. This correction was derived independently for each image. Each image was then reprojected onto a Carrington heliographic grid in cylindrical equal-area projection \cite{Calabretta2002AA,Thompson2006AA}, identical to that of the GONG synoptic magnetogram \cite{Harvey1996Science} for the same rotation: $360\times180$ pixels, uniformly spaced in longitude ($1^{\circ}$) and in the sine of latitude ($\Delta\sin\varphi=1/90$). The resampling used the adaptive, anti-aliased algorithm of DeForest \cite{DeForest2004SoPh} as implemented in the \textit{reproject} package (and output pixels with a reprojection footprint below 0.5 were discarded). The synoptic map was formed as a weighted mean of the reprojected images, $I(\lambda,\varphi)=\sum_{i}w_{i}I_{i}(\lambda,\varphi)\big/\sum_{i}w_{i}$. The weights are $w_{i}=\exp[-\Delta_{i}^{2}/(2\sigma^{2})]$ for $\lvert\Delta_{i}\rvert\leq2.5\sigma$ and zero otherwise, where $\Delta_{i}$ is the difference between the Carrington longitude $\lambda$ and the central-meridian longitude of image $i$. We used $\sigma=3.3^{\circ}$, set by the longitude advance between consecutive images, so that each pixel combines images obtained within $\pm0.6$ days of its central-meridian passage. $\Delta_{i}$ was measured in rotation phase rather than modulo $360^{\circ}$, so that each longitude includes only data from its own central-meridian passage during the CR. Images obtained within $2.5\sigma$ of central-meridian longitude before and after the rotation were also included, so that longitudes near the rotation boundaries were treated in the same way as the rest of the map.

In each case, we place a box approximately on the Carrington maps corresponding to the region in the corona where we believe the emission observed by DKIST/Cryo-NIRSP originates. We note that the dominant contribution to emission at the limb is expected to originate somewhere between $\pm20^{\circ}$ either side of the point directly above the limb \cite[e.g.,][]{Judge2007ApJ,Morton2026RNAAS}. Hence the box indicates where this region is once projected onto the photosphere. The box height is determined by the range of latitude values covered by the slit. We use the header information on the telescope pointing provided by the DKIST Science team but note that this may be inaccurate by up to 10\arcsec.

In general, for the two DKIST/Cryo-NIRSP data sets used in this analysis, it is challenging to determine clearly if the observed emission originates exclusively from an open field region.

\subsection*{Data set 23-03-2025 (CR2295)}
The synoptic AIA map is shown in Supplementary Figure~\ref{sf:aia295}. The overplotted contours indicate the positive and negative patches of open flux -- estimated for $R_{ss}=1.6\,\Rsun$. For increasing values of the source surface radius, the thin strip of open flux near the region identified with the Cryo-NIRSP observation disappears. And for $R_{ss}\approx2.0\,\Rsun$, the source surface polarity inversion line ($\mathrm{PIL}_{\mathrm{SS}}$) changes with the loop to the northern hemisphere around 170--180$^{\circ}$ disappearing. Most of the other open flux features remain largely unchanged.

\begin{sifigure}[t]
\figbox[0 6cm 0 5cm]{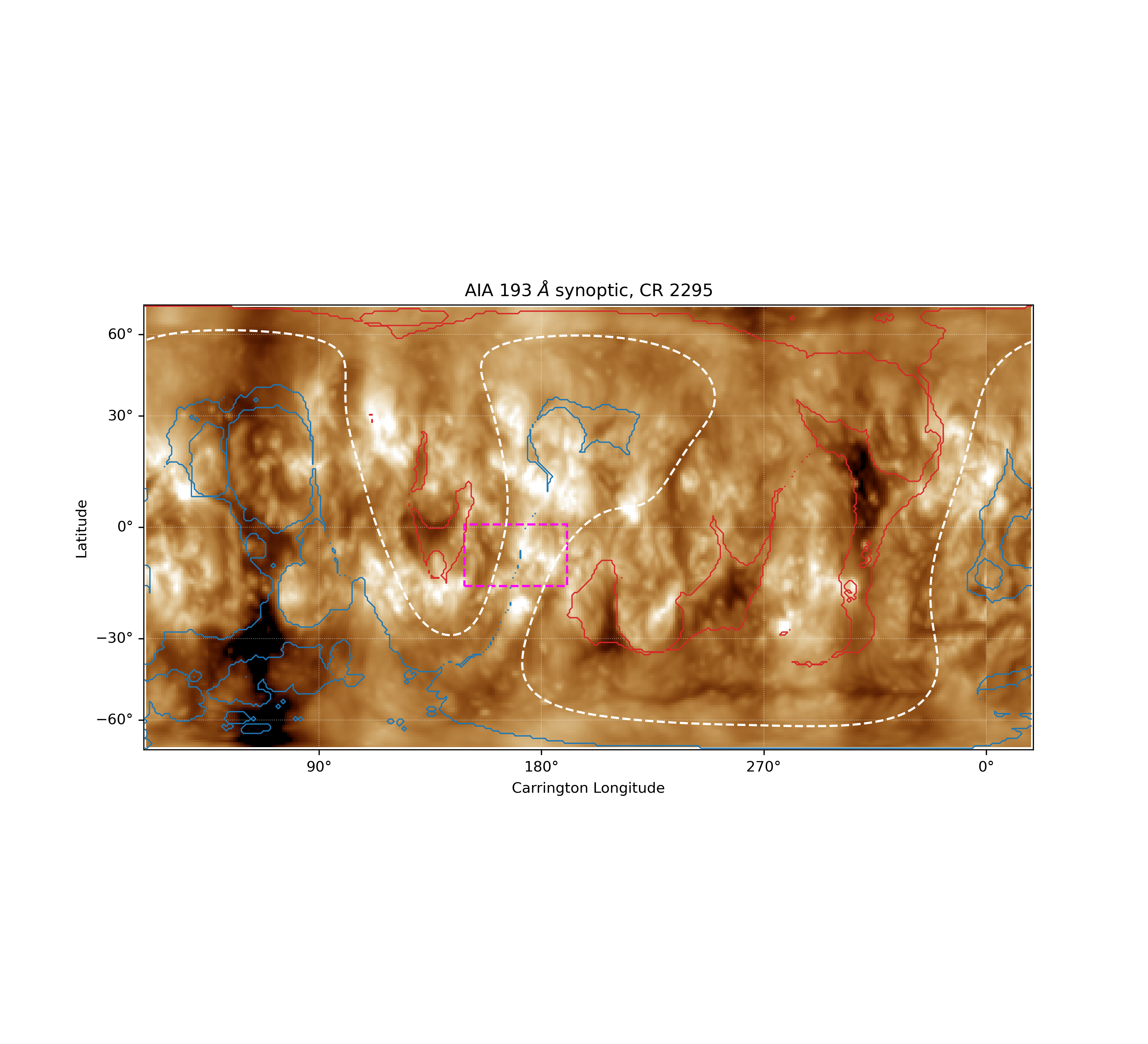}{9cm}{Synoptic AIA 19.3\,nm map, CR2295, with open-flux contours (red/blue), source-surface polarity inversion line (white dashed) and projected Cryo-NIRSP slit box (pink dashed).}
\caption{Synoptic AIA 19.3\,nm emission map for Carrington rotation 2295 with magnetic field features overplotted. The red and blue contours show negative and positive open field regions, respectively. The white dashed line shows the polarity inversion line at the source surface. The pink dashed box is the projected Cryo-NIRSP slit location (covering the longitudinal region expected to contribute to the emission).}
\label{sf:aia295}
\end{sifigure}

Supplementary Figure~\ref{sf:pfss295} shows the magnetic field lines from the potential field extrapolation (with $R_{ss}=1.6\,\Rsun$). The field line foot points are calculated for a local region centred on the boxed pink box shown in Supplementary Figure~\ref{sf:aia295}. We include neighbouring regions to the box -- effectively sampling from a region that is larger than the width and height of the box by $5^{\circ}$ on all sides. The local region near the Cryo-NIRSP slit contains two active regions, which are likely not properly represented by a potential field extrapolation. Further, a filament was present on disk between the two regions, approximately where the wine glass shaped feature is present (Extended Data Figure~\ref{ed:cryo}). The filament erupted the day before the Cryo-NIRSP observations, suggesting the magnetic field in this region substantially changed from when the synoptic magnetogram data were taken. An enhanced version (WOW \cite{Auchere2023AA}) of the AIA image shows the visible fine structure and the area between the two active regions contains little evidence of the loop structures observed in the extrapolation.

The Doppler velocity data that we analysed from Cryo-NIRSP are from the lower portion of the observational slit (pink dashed line), from (1065\arcsec, $-$199\arcsec) to (1057\arcsec, $-$261\arcsec) in helioprojective Cartesian coordinates.
This overlaps with the positive open field from the magnetic field extrapolation. In the enhanced AIA image, there is a corresponding intensity structure that is qualitatively similar in shape to the predicted open field. The data from the portion of the slit used in the analysis may also contain signal originating from the apparently closed, large coronal loop system that is adjacent to the identified open flux region.

\begin{sifigure}[t]
\figbox[0 0 0 0]{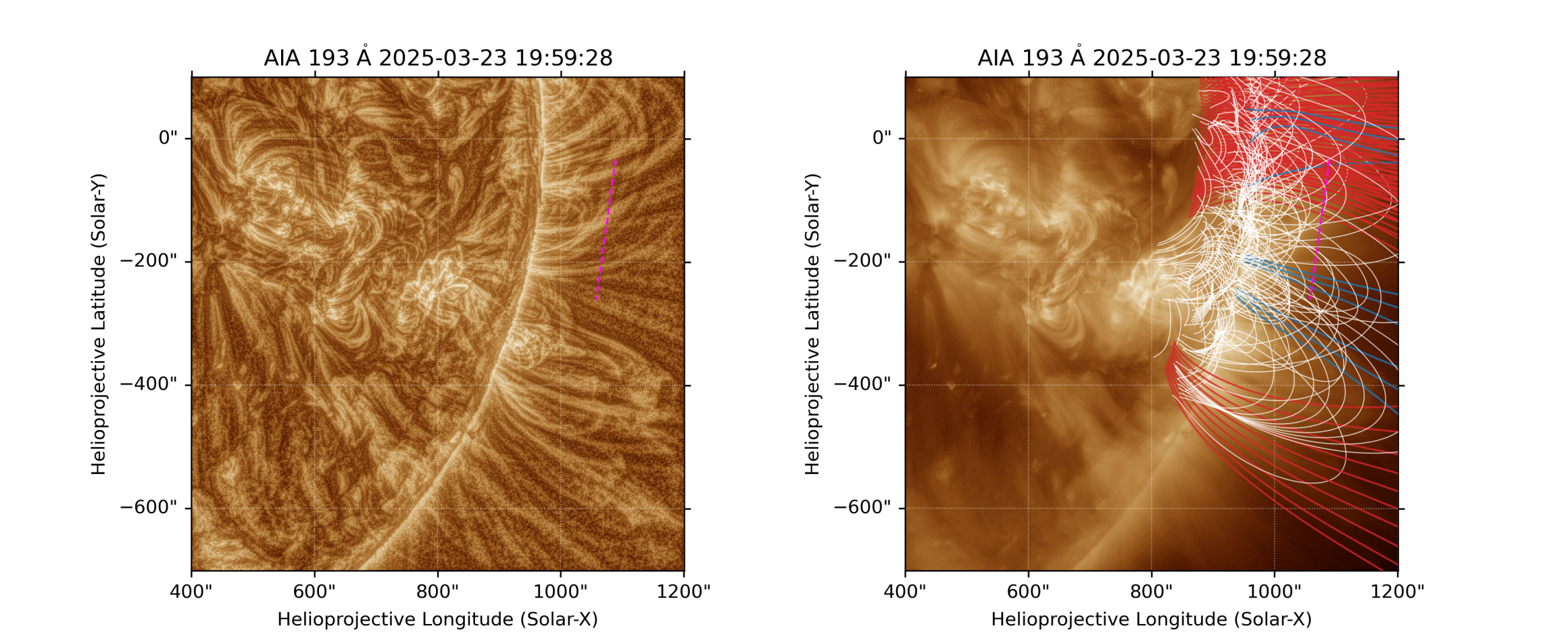}{8cm}{Left: WOW-filtered AIA image. Pink dashed line: Cryo-NIRSP slit.  Right: potential-field lines (open red/blue, closed white) on AIA 19.3\,nm.}
\caption{Left panel is the WOW filtered image (AIA 19.3\,nm) which improves visualisation of the local structures. The pink dashed line is the location of the Cryo-NIRSP slit taken from the header files. Right panel shows the magnetic fields lines associated with the extrapolation overplotted on image of the corona. The red and blue magnetic field lines are open flux, the white are closed magnetic field lines. The right panel }
\label{sf:pfss295}
\end{sifigure}

\subsection*{Data set 30-10-2023 (CR2277)}
The synoptic AIA map for this data set is shown in Supplementary Figure~\ref{sf:aia295}. The Cryo-NIRSP field of view overlaps with a faint dark region in EUV which could be a coronal hole. The potential field extrapolations (with $R_ss=1.6 R_\odot$) indicate there is some open flux open magnetic field found within the projected region (pink box) thought to contribute emission observed with Cryo-NIRSP. The open flux is rooted behind the active region but expands with height and overlaps with the location of the Cryo-NIRSP slit (Supplementary Figure~\ref{sf:pfss277}).

\begin{sifigure}[t]
\figbox[0 12cm 0 2cm]{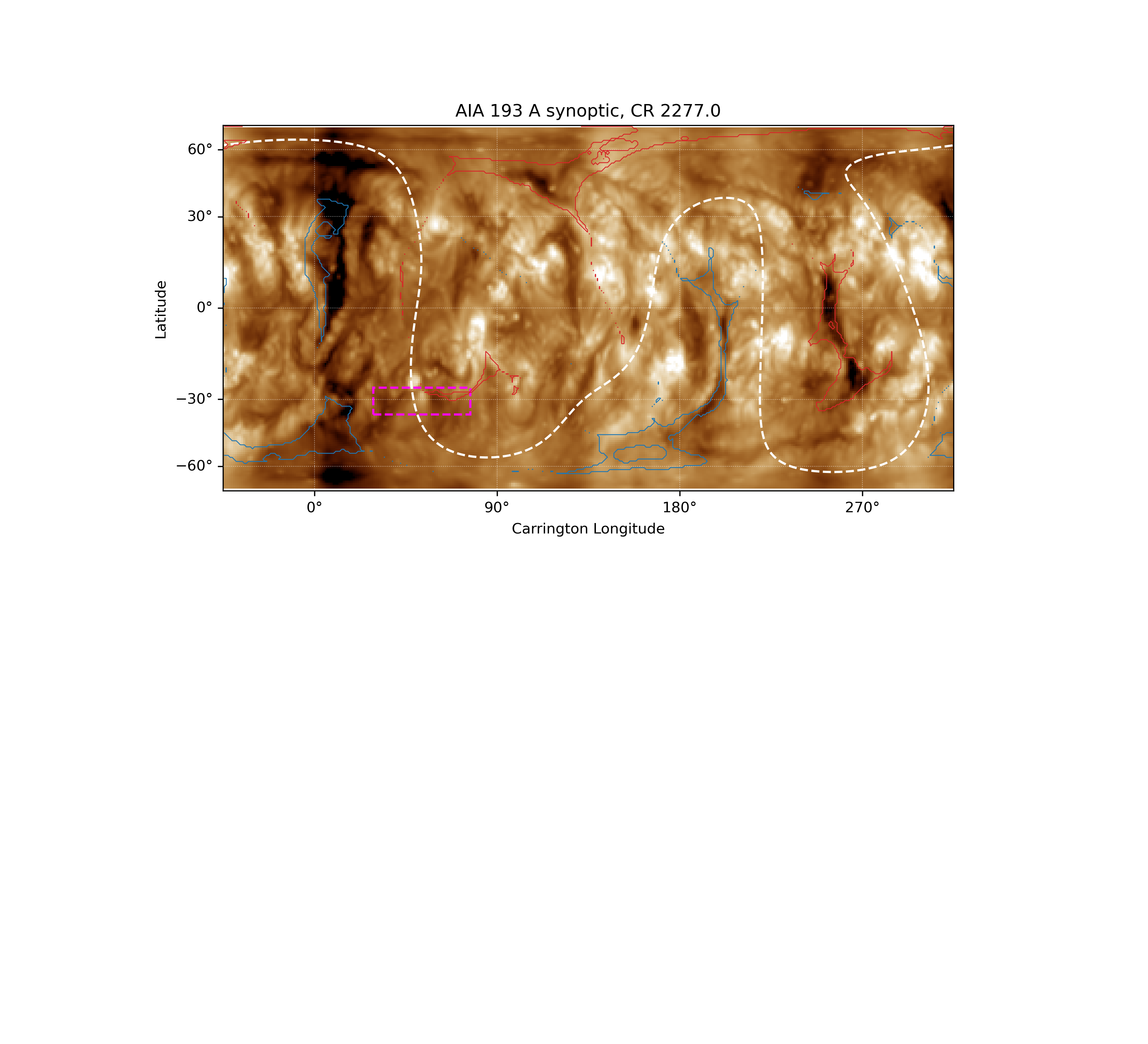}{9cm}{Synoptic AIA 19.3\,nm map, CR2277 (same layout as Supplementary Figure~\ref{sf:aia295}).}
\caption{Same as Supplementary Figure~\ref{sf:aia295} except for Carrington rotation CR2277.}
\label{sf:aia277}
\end{sifigure}

\begin{sifigure}[t]
\figbox[0 2cm 0 2cm]{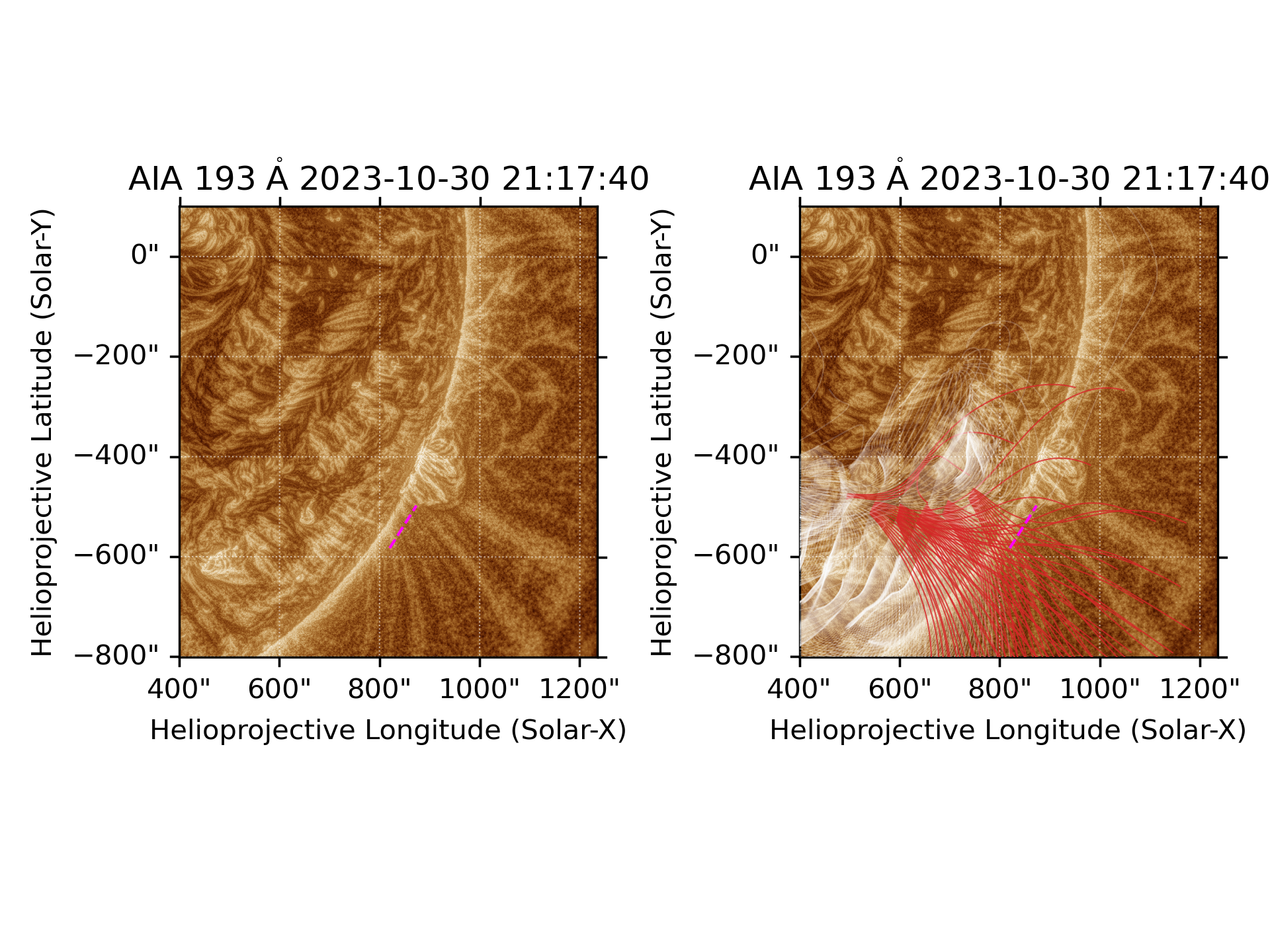}{8cm}{Left: WOW-filtered AIA image. Pink dashed line: Cryo-NIRSP slit.  Right: potential-field lines (open red, closed white) on WOW-filtered AIA.}
\caption{Left panel is the WOW filtered image (AIA 19.3\,nm) which improves visualisation of the local structures. The pink dashed line is the location of the Cryo-NIRSP slit taken from the header files. Right panel shows the magnetic fields lines associated with the extrapolation overplotted on image of the corona . The red and blue magnetic field lines are open flux, the white are closed magnetic field lines.}
\label{sf:pfss277}
\end{sifigure}
\FloatBarrier

\end{document}